\documentclass[journal,twoside,web]{ieeecolor2}
\usepackage{generic}
\usepackage{cite}
\usepackage{amsmath,amssymb,amsfonts}
\usepackage{graphicx}
\usepackage{textcomp}
\usepackage{caption}
\usepackage{enumerate}
\usepackage{flushend}
\usepackage{color,soul}
\usepackage[T1]{fontenc}
\usepackage[font=small,labelfont=bf,tableposition=top]{caption}
\usepackage{afterpage}

\usepackage{cite}
\usepackage{amsmath,amssymb,amsfonts}
\usepackage{multirow}

\usepackage{csquotes}

\usepackage{algorithm}
\usepackage{algpseudocode}
\usepackage{tablefootnote}
\usepackage{footnote}
\usepackage{subfigure}
\usepackage{caption}
\usepackage{graphicx}
\usepackage{textcomp}
\usepackage{hyperref}
\usepackage{color,soul}
\usepackage{tabularx}
\usepackage{ragged2e}
\usepackage{balance}
\usepackage{booktabs}

\newenvironment{compactquote}
  {\begin{list}{}{%
     \setlength{\leftmargin}{1em}%
     \setlength{\rightmargin}{1em}%
   }\item\relax}
  {\end{list}}

\newif\ifhighlight
\highlightfalse
\ifhighlight
  \newcommand{\rev}[1]{\textcolor{blue}{#1}}
  \newcommand{\todoflag}[1]{\textcolor{red}{[#1]}}
\else
  \newcommand{\rev}[1]{#1}
  \newcommand{\todoflag}[1]{}
\fi

\def\BibTeX{{\rm B\kern-.05em{\sc i\kern-.025em b}\kern-.08em
    T\kern-.1667em\lower.7ex\hbox{E}\kern-.125emX}}

\newcommand{\bftext}[1]{\text{\fontencoding{T1}\fontseries{b}\selectfont #1}}

\def\BibTeX{{\rm B\kern-.05em{\sc i\kern-.025em b}\kern-.08em
    T\kern-.1667em\lower.7ex\hbox{E}\kern-.125emX}}
\begin{document}
\title{Extracting Physiological Numeric Values from French Pediatric ICU Notes: A Multi-Objective Representation Learning Approach}
\author{Boammani Aser Lompo*, Thanh-Dung Le*, \IEEEmembership{Senior Member, IEEE}, \\ Philippe Jouvet, and Rita Noumeir, \IEEEmembership{Member, IEEE}
\thanks{*Boammani Aser Lompo and Thanh-Dung Le contributed equally to this work, and should be regarded as co-first authors.}
\thanks{This work was supported in part by the Natural Sciences and Engineering Research Council (NSERC), in part by the Fonds de la recherche en sante du Quebec (FRQS).}
\thanks{Boammani Aser Lompo is with the Biomedical Information Processing Lab, \'{E}cole de Technologie Sup\'{e}rieure, University of Qu\'{e}bec, Canada (email:boammani.lompo.1@ens.etsmtl.ca). }
\thanks{Thanh-Dung Le is with the Biomedical Information Processing Lab, \'{E}cole de Technologie Sup\'{e}rieure, University of Qu\'{e}bec, Canada, and with the Conrad Blucher Institute, Texas A\&M University-Corpus Christi, Texas, USA. (e-mail: thanh-dung.le@tamucc.edu). }
\thanks{Philippe Jouvet is with the CHU Sainte-Justine Research Center, CHU Sainte-Justine Hospital, University of Montreal, Montréal, Québec, Canada. (email: philippe.jouvet.med@ssss.gouv.qc.ca) }
\thanks{Rita Noumeir is with the Biomedical Information Processing Lab, \'{E}cole de Technologie Sup\'{e}rieure, University of Qu\'{e}bec, Canada (email: rita.noumeir@etsmtl.ca) }}

\maketitle
\vspace{-5mm}

\begin{abstract}
\rev{Numeric values in clinical narratives, such as heart rate, oxygen saturation, and pressure gradients, carry diagnostic meaning that Transformer models trained on generic text do not capture. \textit{Objective:} We categorize numerical values in French pediatric intensive care unit (PICU) notes into eight physiological categories using CamemBERT-bio, under two constraints that make large-scale LLMs impractical: only 1,072 real, annotated clinical samples are available for this rare, single-site condition, and training must run on GPUs shared concurrently with other hospital workloads rather than a dedicated cluster. \textit{Methods:} We compare fine-tuning CamemBERT-bio with Label Embedding for Self-Attention (LESA) against combining LESA with Xval, a magnitude-aware number embedding, under a multi-objective training loss. \textit{Results:} Standard fine-tuning did not improve $F_1$ score, but CamemBERT-bio + LESA raised it by over 13\%, and adding Xval matched this gain while approaching GPT-4's performance. \textit{Conclusion:} LESA and Xval let a compact encoder achieve reliable physiological value extraction under limited real data and shared hospital compute, offering a practical alternative to large-scale LLMs. \textit{Significance:} Under limited-data and shared-compute constraints, this compact BERT-based language model remains effective without the resource trade-offs of trillion-parameter LLMs.}

\end{abstract}


\section{Introduction}
\IEEEPARstart{A}{} major challenge in processing medical texts is understanding numerical data. Numbers are ubiquitous in medical documents, appearing as physiological measurements, event occurrences, durations, and protocol codes (e.g., G1P2 for Gravida 1 Para 2). In Clinical Decision Support Systems (CDSS), recognizing and interpreting these numbers within their specific context is crucial to determine whether they indicate a problem: in \textit{``Patient agé de 12 ans, rythme cardiaque de 130 bpm,"} "12" must be identified as the patient's age and "130" as the heart rate, and a heart rate of "130" understood as problematic for a 12-year-old.

\rev{Three constraints compound in our setting: a small, imbalanced annotated corpus, since data annotation in healthcare is costly and medical notes are unstructured and full of jargon and incomplete sentences; a wide numeric range, values in our dataset span $10^{-4}$ to $10^5$, so a single large outlier can destabilize training; and training GPUs that are a shared hospital asset used concurrently by other clinical projects rather than a dedicated cluster. Together, these constraints rule out the large-scale LLMs that dominate current numerical-reasoning work and motivate a compact encoder instead.}

This paper contributes to a broader project aimed at leveraging a LLM for diagnosing cardiac failure, which the National Heart, Lung, and Blood Institute (2023) defines as the heart's inability to pump sufficient blood to meet the body's needs, presenting through indicators such as blood pressure, ventricular gradients, and ejection fractions. \rev{Our research extracts physiological numeric values from French pediatric intensive care unit (PICU) notes by classifying} numerical values extracted from medical notes into one of eight specific physiological categories using CamemBERT-bio \cite{touchent2023camembert}.

\vspace{-3mm}

\subsection{Goal statement}
Our objective is to train CamemBERT-bio on a limited and uneven dataset\rev{, under the shared-compute constraints,} with the aim of accomplishing the following outcome:

\noindent \textbf{Input medical note}:
\begin{compactquote}
\textit{``BB né à 37+6 par AVS, APGAR 8-8-8. Dx Anténatale: d-TGV avec septum intact, sat 50-65\% à la naissance."}
\end{compactquote}
\noindent \textbf{Expected output}: as shown in the following Table \ref{tab1}.

\begin{center}
\captionof{table}{\rev{Expected output for the example clinical note.}}
\label{tab1}
\begin{tabular}{@{}lll@{}}
\toprule
\textbf{Value} & \textbf{Attribute} & \textbf{Unit} \\
\midrule
$37+6$ & Divers (âge du patient) & semaines \\
$8-8-8$ & Score APGAR & \\
$50-65$ & Saturation en oxygène & \% \\
\bottomrule
\end{tabular}
\end{center}

\subsection{Contribution} 
In this work, we outline the subsequent contributions:
\begin{itemize}
\item We demonstrate that CamemBERT-bio + LESA as introduced in \cite{lompo2025mediumsizedtransformersmodelsrelevant} can be further enhanced through language modeling prefinetuning. This prefinetuning does not improve CamemBERT-bio alone, indicating that integrating the Label Embedding for Self-Attention (LESA) technique necessitates pretraining to adjust the model's parameters. Pre-finetuning on a small medical dataset significantly improves the $F_1$ score for the resulting CamemBERT-bio + LESA model.
\item We develop a new model by extending finetuned CamemBERT-bio + LESA with the Xval architecture \cite{golkar2023xval}. The main idea of Xval is to incorporate the magnitude of numbers into their embeddings before the attention layers. This new model provides a multi-objective representation of numbers, enhancing generalizability and robustness across classification categories.
\item We assess and juxtapose our tailored model against conventional counterparts {\color{black}including LLMs such as GPT-4}. Findings demonstrate that our model surpasses standard approaches and adeptly manages datasets characterized by limited size and imbalance. {\color{black}Additionally, our model delivers performance on par with GPT-4 despite the significant disparity in parameter size.}
\rev{\item We replace the learned loss weighting of \cite{kendall2018multi} with a log-scaled number-prediction loss $\tilde{L}_2$ under fixed equal weights. Section \ref{sec:novelty} shows analytically that the learned weighting collapses onto the contextual objective whenever the target range is wide, and that the log-scaled gradient is bounded by the predicted magnitude. This is the component that makes a magnitude objective usable on clinical values spanning $10^{-4}$ to $10^{5}$.}
\rev{\item We report a per-class Welch $t$-test over ten seeds for every baseline comparison, which pins the benefit of the multi-objective representation to specific classes instead of inferring it from mean scores alone.}
\end{itemize}

\rev{\noindent The remainder of this paper is organized as follows. Section \ref{related} reviews prior work on numerical representation and identifies the specific shortcomings this study addresses, summarised in Table \ref{tab:priorwork}. Section \ref{task and dataset} defines the classification task and describes the clinical corpus. Section \ref{sec:limitations-e2e} documents the two properties of clinical notes, tokenization and missing sentence structure, that motivate the design. Section \ref{approach} presents the methodology, including the notation of Table \ref{tab:notation}, the novelty statement of Section \ref{sec:novelty} and the training procedure of Algorithm \ref{alg:model2}. Section \ref{sec:experiments} details the baselines and the hyperparameters of Table \ref{tab:hyperparams}. Section \ref{sec:results} reports the results and the significance analysis, Section \ref{sec:limits} states the limitations, and Section \ref{sec:conclusion} concludes.}

\begin{figure*}[!t]
\centering
\includegraphics[width=0.9\textwidth]{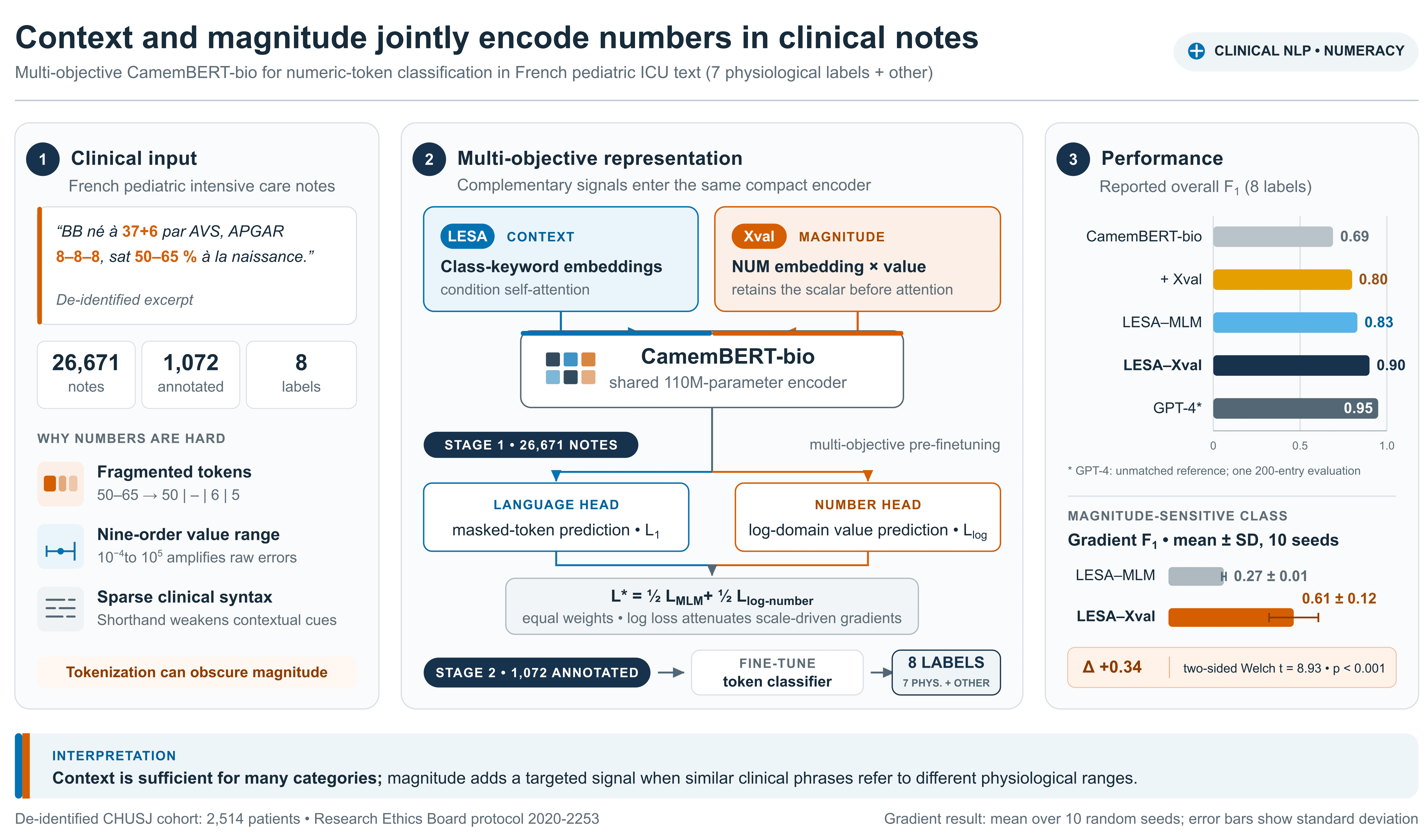}
\caption{\rev{Study overview. \textbf{(1)} Numeric tokens in French pediatric intensive care notes are classified into eight physiological categories despite fragmented tokenization, a nine-order-of-magnitude value range, and limited sentence structure. \textbf{(2)} The proposed model combines class-aware self-attention (LESA) with magnitude-aware token embeddings (Xval) in a shared CamemBERT-bio encoder trained with equally weighted contextual and log-scaled number-prediction objectives. \textbf{(3)} The model achieves an overall $F_1$ of $0.90$ with 110 million parameters,  compared with $0.95$ for GPT-4, and improves the magnitude-dependent \textit{Gradient} class from $0.27$ to $0.61$.}}
\label{fig:overview}
\vspace{-3mm}
\end{figure*}

\rev{Figure \ref{fig:overview} presents the argument of the paper in a single view: the properties of the corpus that defeat a standard encoder, the two representations introduced to address them, and the performance that results. We place it at the outset because the contribution of this work is a training objective, not a new architecture, and an objective is only meaningful in relation to the failure it is designed against. Reading the three parts together, a specific defect of the data, the loss constructed in response, and the one class on which that loss pays off, conveys the logic of the study more directly than the same material distributed across three separate sections.}
\vspace{-3mm}

\section{Related works}\label{related}

\rev{Processing numerical values with LLMs is a highly active research area. Although \cite{wallace2019nlp} claimed that most word embedding models capture numeracy, several limitations persist. \cite{cui2019regular} used a rules-based method with regular expressions, whose accuracy is contingent on clean text and consistent abbreviations; \cite{chen2023improving} proposed a question-answer pre-training task; and other work explored alternative number tokenization, digit-by-digit, scientific notation, or customized schemes \cite{charton2021linear}. A number-agnostic line of work instead replaces numbers with placeholder keywords and scales their embeddings for context-centric encoding \cite{golkar2023xval, Loukas_2022, lompo2025mediumsizedtransformersmodelsrelevant}. \cite{lompo2025mediumsizedtransformersmodelsrelevant} introduced Label Embedding for Self-Attention (LESA), incorporating class descriptions into the initial feature extraction phase; it is similar to EmbedNum \cite{Nguyen2018EmbNumSL}, which classifies numbers by comparing their embeddings to category embeddings, but LESA uses that similarity only to construct the number's embedding, not for classification itself. Advanced LLMs such as LLaMA, GPT-4, and Claude, by contrast, adjust to diverse contexts autonomously and can handle numerical values without preprocessing.}

\rev{Despite their strong performance, LLMs face limitations that motivate a compact-encoder alternative for our setting. Model size alone does not resolve complex semantic understanding \cite{cheng2024potential}; \cite{shah2024accuracy} found GPT-3.5 efficient at extracting clinical information but prone to abbreviation errors, and \cite{ullah2024challenges} reports contextual, interpretability, and training-bias issues for GPT-4 on medical applications. Deploying such models also raises patient-data privacy concerns, since they retain vast amounts of training data in their parameters. Given these challenges and their computational cost, we prioritize a smaller, more adaptable encoder, CamemBERT-bio, for medical NLP.}

Another significant challenge is handling small and imbalanced datasets. While transformer-based architectures generally perform well with large datasets, research by \cite{Li2018, ezen2020comparison, mascio2020comparative} indicates that they may not outperform traditional methods, such as LSTM, when dealing with smaller datasets of clinical notes. A common technique to enhance language models' performance on small datasets is Knowledge Distillation (KD) \cite{Bucila2006ModelC}, which trains a smaller student network to replicate a larger teacher's behavior and is the key idea behind DistilBERT \cite{Sanh2019} and its French counterpart, DistilCamemBERT \cite{delestre2022distilcamembert}. Another approach to handling limited data is to construct multiple representations for words as a weighted sum of losses, each specialized in a different aspect of the modality; this raises the issue of weighting each loss appropriately, which \cite{kendall2018multi} addresses by an automatic method optimizing the log-likelihood of the model predictions on vision targets, a result we found relevant to our case.

\rev{Table \ref{tab:priorwork} consolidates this discussion into an explicit statement of what each family of methods leaves unresolved, since it is the union of these gaps that defines the design space for the present work. Two observations follow from the table. First, the existing techniques complement each other more than they compete: placeholder-based methods supply magnitude but discard the label semantics, whereas label-embedding methods supply class semantics but remain blind to magnitude. No prior study combines the two in a single encoder. Second, the one mechanism that would allow such a combination, namely an automatic weighting of heterogeneous losses, was derived for targets of comparable scale and has not been examined when one target spans nine orders of magnitude. Section \ref{sec:novelty} closes that gap, and the remainder of the paper evaluates the resulting model.}
\vspace{-3mm}

\begin{table*}[!t]
\caption{\rev{Summary of representative approaches to numerical representation and the limitation each one leaves unresolved.}}
\label{tab:priorwork}
\centering
\rev{%
\begin{tabular}{@{}p{0.20\textwidth}p{0.20\textwidth}p{0.50\textwidth}@{}}
\toprule
\textbf{Family of methods} & \textbf{Representative work} & \textbf{Limitation left unresolved} \\
\midrule
Rule-based extraction & Cui \textit{et al.} \cite{cui2019regular} & Accuracy is contingent on clean text and consistent abbreviations, both of which fail on free-text clinical notes. \\
\addlinespace
Auxiliary QA pre-training & Chen \textit{et al.} \cite{chen2023improving} & Requires a large auxiliary question--answer corpus that does not exist for French pediatric intensive care. \\
\addlinespace
Alternative tokenization & Charton \cite{charton2021linear}, Zhang \textit{et al.} \cite{zhang2020language} & Digit-level and scientific-notation schemes still fragment a value across sub-tokens, so magnitude remains implicit. \\
\addlinespace
Placeholder and scaling & Golkar \textit{et al.} \cite{golkar2023xval}, Loukas \textit{et al.} \cite{Loukas_2022} & Magnitude is injected into the embedding, but the accompanying regression loss is unstable when values span many orders of magnitude. \\
\addlinespace
Label embedding & Lompo \textit{et al.} \cite{lompo2025mediumsizedtransformersmodelsrelevant}, Nguyen \textit{et al.} \cite{Nguyen2018EmbNumSL} & Conditions the encoder on class semantics but carries no magnitude information, so numerically separable classes stay confused. \\
\addlinespace
Knowledge distillation & Sanh \textit{et al.} \cite{Sanh2019}, Delestre and Amar \cite{delestre2022distilcamembert} & Reduces parameter count under a small-data regime without addressing numeracy in any form. \\
\addlinespace
Learned multi-loss weighting & Kendall \textit{et al.} \cite{kendall2018multi} & Derived for vision targets of comparable scale; the weighting collapses onto the dominant objective when the ranges diverge, as shown in Section \ref{sec:novelty}. \\
\addlinespace
Large-scale LLMs & Hu \textit{et al.} \cite{hu2024improving}, Shah \textit{et al.} \cite{shah2024accuracy}, Ullah \textit{et al.} \cite{ullah2024challenges} & Strong accuracy, but computational cost, abbreviation errors and the transfer of patient data to external services restrict deployment in a hospital. \\
\bottomrule
\end{tabular}}
\end{table*}

\section{The task and the dataset}\label{task and dataset}

\rev{The dataset was supplied by the Pediatric Intensive Care Unit at CHU Sainte-Justine (CHUSJ) and includes children under 18: 26,671 unannotated medical notes from \textbf{2,514 de-identified patients}, of which only 1,072 samples ($\approx$30,000 tokens) are annotated. These notes were selected as they capture the physician's reasoning behind a patient's admission and the initial care plan. Access was obtained following approval of our research protocol (\#2020-2253) by the CHUSJ Research Ethics Board.}

\noindent Our objective is to systematically classify tokens into eight categories, each corresponding to a key medical parameter, to enhance the identification of heart failure risks from the patient's medical history:

\begin{itemize}
\item[$\circ$] \textbf{Contractilité} (contractility): This category encompasses ejection fraction and shortening fraction, which are indicative of the heart’s ability to contract. It is represented as \textbf{Cp}.
\item[$\circ$] \textbf{Fréquence cardiaque} (heart rate): This will be referred to as \textbf{FC}.
\item[$\circ$] \textbf{Diamètre de l’artère pulmonaire} (pulmonary artery diameter): Denoted as \textbf{D}.
\item[$\circ$] \textbf{Saturation en oxygène} (oxygen saturation): Noted as \textbf{SO2}.
\item[$\circ$] \textbf{APGAR} (APGAR score): It will be abbreviated as \textbf{APGAR}.
\item[$\circ$] \textbf{Gradient} (pressure gradient): This represents the difference in blood pressure between the heart’s ventricles, without considering directional flow. It is labeled as \textbf{G}.
\item[$\circ$] \textbf{CIA-CIV}: Size of the atrial or ventricular septal defect. These measurements are crucial for assessing the severity of congenital heart malformations. This category is denoted as \textbf{CIA-CIV}.
\item[$\circ$] \textbf{Out of class}: Items that do not fit into the previously defined categories. Labeled as \textbf{O}.
\end{itemize}

\begin{figure}
\centering
\begin{subfigure}
  \centering
  \includegraphics[width=0.86\linewidth]{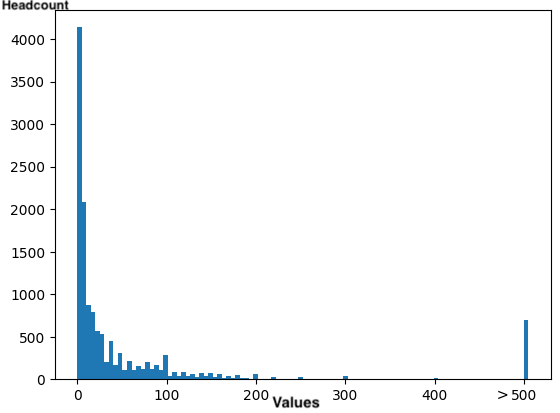}
\end{subfigure}%
\begin{subfigure}
  \centering
  \includegraphics[width=0.86\linewidth]{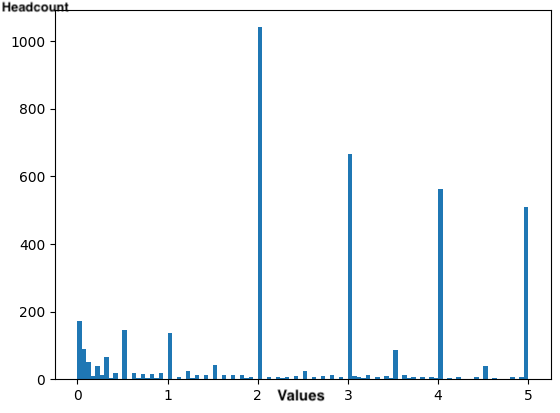}
\end{subfigure}
\caption[Number distribution]{(a) Overview of the whole number distribution (b) A focus on the small number distribution. Non-integer numbers are effectively represented in the dataset.}
\label{fig:number distribution}
\vspace{-3mm}
\end{figure}

The number distribution in the dataset is displayed in Figure ~\ref{fig:number distribution}. All the numbers beyond $500$ are regrouped because they are scattered from $500$ to $35000$.

\section{A few limitations of end-to-end learning for Transformers Models in medical datasets}\label{sec:limitations-e2e}
\rev{Most pretrained Transformer models, such as GPT-4, LLaMA, and BERT, are trained end-to-end, from raw input to final output without intermediate steps or feature engineering. Unlike many of these state-of-the-art approaches, our work faces the added constraint of a small, imbalanced dataset. This section covers the other major limitations of end-to-end training for Transformers specifically encountered in the medical domain.}
\vspace{-3mm}

\subsection{The Limitations of Tokenization}\label{tokenization problem}

\rev{Recent transformer-based models rely on efficient subword tokenizers such as WordPiece \cite{wu2016google} and SentencePiece \cite{kudo-richardson-2018-sentencepiece}, which decompose infrequent words into meaningful character subsequences, e.g., \textbf{``working"} $\to$ \textbf{``work"} + \textbf{``ing"}. As noted by \cite{Loukas_2022}, this breaks down for numerical values: \textbf{``123.7g/L"} is disassembled into \textbf{``12"} + \textbf{``3"} + \textbf{"."} + \textbf{``7"} + \textbf{``g/L"}. Since only the first token of a segmented word is typically classified, only the ambiguous \textbf{``12"} would be seen, and the magnitude difference between \textbf{``12"} and \textbf{``123.7"} is lost. We claim that self-attention and positional embeddings alone are insufficient to recover this magnitude information reliably.}

\begin{figure}
\centering
\includegraphics[width=1\linewidth]{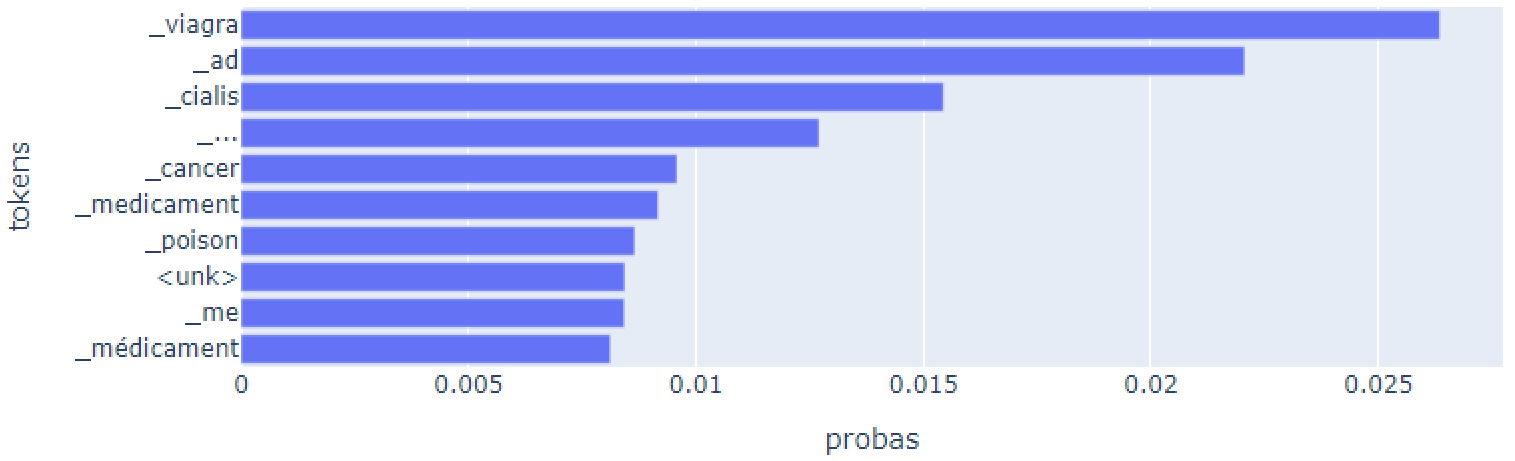}
  \caption{Output of a completion task by CamemBERT-bio. The input sentence is \textit{``Patient en détresse respiratoire, gradient VG-VD ad $\langle$ mask $\rangle$ mmgh."}. The figure contains the possible words to fill the mask with their corresponding predicted probabilities.}
  \label{completion with camembert bio}
  \vspace{-3mm}
\end{figure}

To illustrate this point, let's examine CamemBERT-bio's performance in a task focused on completing sentences that contain numerical values (Figure ~\ref{completion with camembert bio}). A word completion task involves taking a sentence and replacing one or more words with the token $\langle$mask$\rangle$. The model is then asked to predict the missing words. The input sentence, \textit{``Patient en détresse respiratoire, gradient VG-VD ad $\langle$ mask $\rangle$ mmgh."}, was crafted to align with the language style of our target dataset. This illustration shows that the model struggles to provide meaningful numerical predictions, as the predicted tokens are not even numbers. 
\vspace{-3mm}
\subsection{The Structure Issue}\label{structure issue}

\rev{Prior work shows sentence structure matters for transformer models: \cite{clark2019does} found that, without explicit supervision, certain BERT attention heads specialize in grammatical relations (direct objects, determiners, prepositional and possessive objects) with over 75\% accuracy. Our medical notes stand apart in this regard, as a significant proportion lack well-defined structure, with clear subjects and verbs sometimes absent altogether, as illustrated below:}\looseness=-1

\begin{itemize}
    \item ``\textit{CARDIOMYOPATHIE  1) S'est présenté le 28/04 à l'hôpital - Admis ad 07/05 aux soins [...] Depuis, céphalée on/off, fatigue, essoufflement.[...] Vomissement le soirx 5 jours.\phantom{eeeee}    Pincement thoracique à 2-3 reprises depuis 1mois. l'a réveillé 1 fois.}"
    \item ``\textit{ATCDp: \#PNA droite à 1 mois de vie et en 2004 \phantom{eeeee}  Atrophie/asymétrie rénale 2aire   Vu en néphro: \phantom{eeeee}   Écho (11/2014): Rein Droit: 8.8 cm, contour supérieur légèrement irrégulier + vague zone kystique, de 5 mm, [...].}"
    \item \textit{``[...] FR 21 FC 100-110 FR 50 [...]."}
\end{itemize}

As noted in \cite{lompo2025mediumsizedtransformersmodelsrelevant}, the last excerpt illustrates the lack of contextual information in sentence structures. This excerpt reads a sequence of physiological indicators. The initial reference to \textit{FR} most likely indicates the respiratory rate (fréquence respiratoire) since a value of 21 is usually too low for the shortening fraction (fraction de raccourcissement). Conversely, the second occurrence of \textit{FR} refers to a value above 50, suggesting it pertains to the shortening fraction. In this case, it was the values themselves, rather than the context, that provided the necessary distinctions, highlighting a flaw in the sentence structure. This issue echoes the challenges of tokenization, emphasizing the need for a thorough understanding of the numerical values. {\color{black} However, it is worth mentioning that the impact of this issue is very limited on advanced LLMs such as LLAMA, GPT-4, and Claude that are engineered to be able to work with data without requiring any preprocessing. Nevertheless, their ability to handle the lack of structure significantly depends on their prompts \cite{hu2024improving}}.

\section{Methodology}\label{approach}
The methodology adopted in this study involves developing a multi-objective representation of numbers to achieve a more generalizable and robust model \cite{collobert2008unified}. Our representations are obtained through the Masked Language Modeling (MLM) Loss and the Number Prediction Loss. \rev{These objectives yield the following two models:}

\begin{itemize}
    \item \rev{CamemBERT-bio + LESA (MLM-adapted), obtained by adapting CamemBERT-bio + LESA to our unannotated medical notes using MLM;}
    \item \rev{CamemBERT-bio + LESA + Xval, the proposed approach trained with both the MLM and number-prediction objectives jointly.}
\end{itemize}

We anticipate that this multi-objective training approach will produce number representations that incorporate both contextual and magnitude information.

\rev{The derivation that follows introduces a number of symbols. Table \ref{tab:notation} collects them in one place, grouped by the stage of the pipeline in which they appear, so readers can look a symbol up in one place instead of hunting for its definition in the surrounding prose.}

\begin{table}[!t]
\caption{\rev{Notation used throughout Section \ref{approach}.}}
\label{tab:notation}
\centering
\rev{%
\footnotesize
\begin{tabular}{@{}ll@{}}
\toprule
\textbf{Symbol} & \textbf{Description} \\
\midrule
\multicolumn{2}{@{}l}{\textit{Input and encoder}} \\
$\boldsymbol{t}$ & Input token sequence of a clinical note \\
$L$ & Number of tokens in $\boldsymbol{t}$ \\
$D$ & Dimension of the embedding space \\
$n$ & Number of classes ($n=8$) \\
$\boldsymbol{x}_{token_i}$ & Initial embedding of the $i$-th token \\
$\boldsymbol{X}$ & Token embedding matrix, $\mathbb{R}^{(L+1)\times D}$ \\
\addlinespace
\multicolumn{2}{@{}l}{\textit{Label embedding for self-attention}} \\
$\boldsymbol{X}^{l}$ & Class-keyword embedding matrix, $\mathbb{R}^{n\times D}$ \\
$\boldsymbol{A}^{l}$ & Cross-attention of $\boldsymbol{X}^{l}$ and $\boldsymbol{X}$, $\mathbb{R}^{n\times (L+1)}$ \\
$\bftext{CoSim}$ & Cosine similarity matrix, $\mathbb{R}^{(L+1)\times (L+1)}$ \\
\addlinespace
\multicolumn{2}{@{}l}{\textit{Magnitude injection}} \\
$t_{masked}$ & Text with every number replaced by \textit{NUM} \\
$h_{masked}$ & Embedding matrix of $t_{masked}$ \\
$h$ & $h_{masked}$ after scaling by the numeric values \\
\addlinespace
\multicolumn{2}{@{}l}{\textit{Objectives and parameters}} \\
$\Theta$ & Full set of model parameters \\
$\theta$ & A single parameter, $\theta \in \Theta$ \\
$\partial_\theta$ & Partial derivative with respect to $\theta$ \\
$f_1(x,\Theta)$ & Logit vector of the language head \\
$f_2(x,\Theta)$ & Scalar output of the number prediction head \\
$\mathcal{D}$ & Training set of triples $(x,y_1,y_2)$ \\
$y_1$ & Index of the masked token in the vocabulary \\
$y_2$ & Value of the masked number, $1$ if not a number \\
$L_1$ & Cross-entropy loss of the MLM objective \\
$L_2$ & Squared-error loss of the number objective \\
$\tilde{L}_2$ & Log-scaled variant of $L_2$, Equation \eqref{eq:logloss} \\
$w_i$ & Weight applied to loss $L_i$ \\
$\sigma_1,\sigma_2$ & Task-uncertainty parameters of \cite{kendall2018multi} \\
$\mathcal{L}(\Theta)$ & Global objective, $\mathbb{E}_{\mathcal{D}}[L(x,y_1,y_2)]$ \\
\addlinespace
\multicolumn{2}{@{}l}{\textit{Significance analysis, Section \ref{sec:results}}} \\
$\bar{x}_k,\, s_k,\, n_k$ & Mean, sample SD and number of runs of model $k$ \\
$u_k,\, t,\, \nu$ & Squared std.\ error, Welch statistic and d.o.f., Equation \eqref{eq:welch-t} \\
$F_{\nu}$ & Student CDF with $\nu$ degrees of freedom \\
\bottomrule
\end{tabular}}
\end{table}

\subsection{Masked Language Modeling (MLM) objective with LESA}\label{mlm}

\rev{We summarize the model description from \cite{lompo2025mediumsizedtransformersmodelsrelevant}. Given a text $\boldsymbol{t} = [token_1, \cdots, token_L]$ of $L$ tokens with initial embeddings $[\boldsymbol{x}_{token_1}, \cdots, \boldsymbol{x}_{token_L}]$, prefixed with the special token $[CLS]$ and fed into the transformer layers:}
\begin{equation}
\boldsymbol{X} = [\boldsymbol{x}_{CLS}, \boldsymbol{x}_{token_1}, \cdots, \boldsymbol{x}_{token_L}] \in \mathbb{R}^{(L+1)\times D}
\end{equation}

\noindent \rev{where $D$ is the embedding dimension. For each class, descriptive keywords are selected and their initial embeddings averaged, giving a label embedding matrix $\boldsymbol{X}^{l}\in \mathbb{R}^{n\times D}$ ($n$ classes). The cross-attention between $\boldsymbol{X}^{l}$ and $\boldsymbol{X}$ is then computed, $\boldsymbol{A}^{l} \in \mathbb{R}^{n\times (L+1)}$, alongside the ordinary self-attention of $\boldsymbol{X}$, $\bftext{Self-attention}\in \mathbb{R}^{(L+1)\times (L+1)}$. With the cosine similarity matrix $\bftext{CoSim} = \textrm{norm}(\boldsymbol{A}^{l})^T \textrm{norm}(\boldsymbol{A}^{l})$, the enhanced self-attention is:}
\begin{equation}
     \bftext{New-Self-attention} = \bftext{Self-attention} + \bftext{CoSim}
\end{equation}

\begin{figure}
\centering
\includegraphics[scale=0.46]{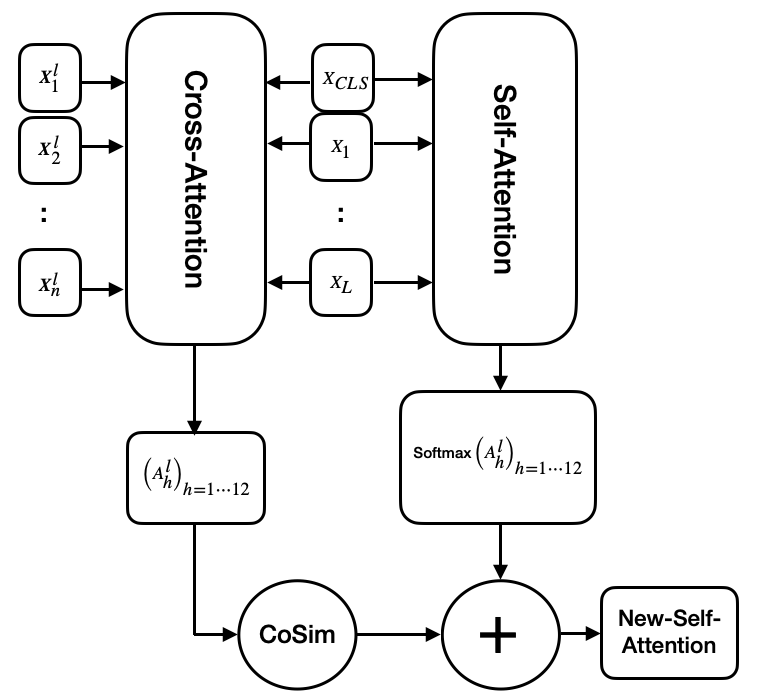}
  \caption{Excerpt from \cite{lompo2025mediumsizedtransformersmodelsrelevant}. An illustration of the Label Embedding for Self-Attention (LESA) is shown. The input to this Self-Attention layer consists of the token embeddings $[X_{CLS}, X_1, \cdots, X_L]$ and the keyword embeddings $[X^l_1, X^l_2, \cdots, X^l_n]$. This layer outputs the resulting enhanced self-attention used by the encoder.}
  \label{fig:lesabert}
  \vspace{-3mm}
\end{figure}

\rev{We use this approach to address the limited size of our dataset, as it reduces overfitting while making the model more robust, then train via the Masked Language Modeling (MLM) task as a prefinetuning step: randomly masking tokens and asking the model to predict them, which builds contextual token representations.}
\vspace{-3mm}

\subsection{Number Prediction Objective}\label{xval}
\rev{This method follows the four steps of \cite{golkar2023xval}. Given input text $t$ containing numbers and words, all numbers are replaced by a placeholder keyword \textit{NUM}, forming $t_{masked}$, with the original values saved. We tokenize and embed $t_{masked}$ into $h_{masked}$, then scale the vectors corresponding to the replaced numbers by their saved values, giving matrix $h$, which is fed to the encoder (Figure \ref{fig:diagram multi}). This placeholder replacement resolves the tokenization limitation noted earlier, and trains the model to learn number magnitude, an important discriminator between classes: heart-rate values, for instance, are usually much larger in magnitude than APGAR values.}
\vspace{-3mm}

\subsection{Multi Objective Loss}

Working with multiple objective losses requires merging them into one single loss
$$L = \sum w_iL_i $$
where $w_i$ represents the weight of the loss $L_i$. {\color{black} In our case we have two losses $L_1$ which corresponds to the MLM objective and $L_2$ which corresponds to the Number Prediction Objective. Our training dataset $\mathcal{D}$ is made of triples $(x,y_1, y_2)$ where $x$ is a sequence of tokens with one token masked using the keyword \textit{Mask}, $y_1$ is the index of the masked token in the model vocabulary, and $y_2$ is the value of the masked token if it was a number. Otherwise $y_2$ is set to $1$. Figure ~\ref{fig:diagram multi} illustrates the inference process of the model. In this illustration, the first masked token is decoded as the standard word "Saturation," and the default number $1$ is assigned to it. The second masked token is decoded as the special token "NUM," and the number prediction head predicts $65$.

\begin{figure}
\centering
\includegraphics[width=1\linewidth]{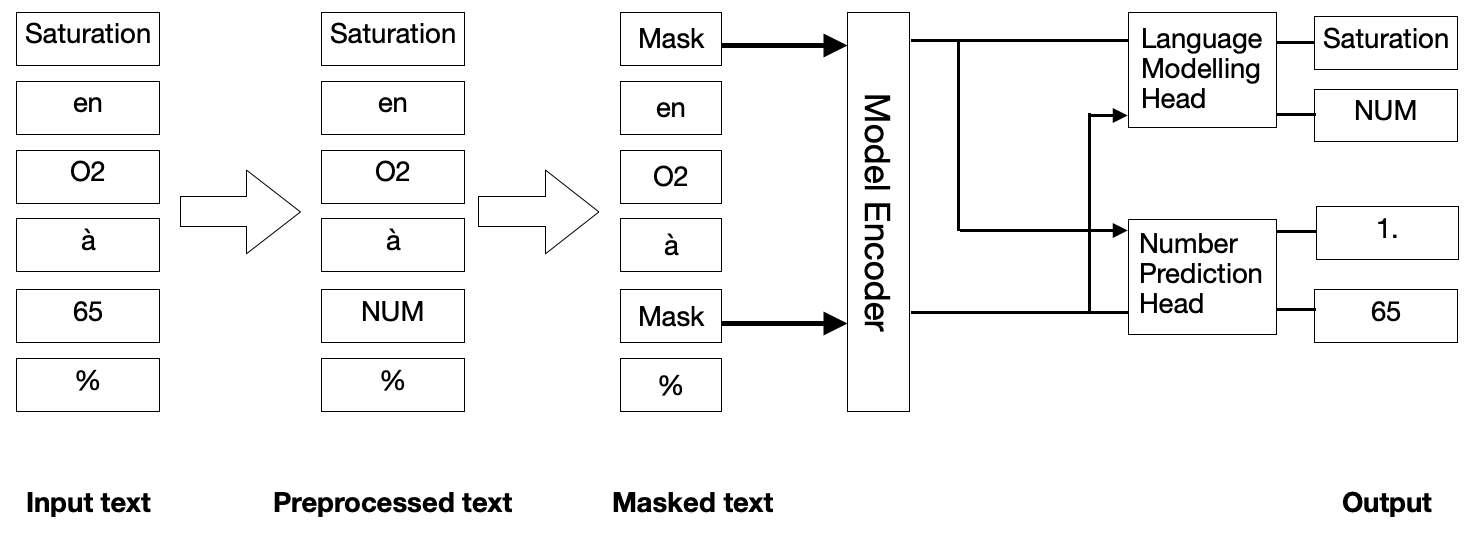}
  \caption{Brief illustration of the inference process using multiple representations.}
  \label{fig:diagram multi}
  \vspace{-3mm}
\end{figure}

In \cite{kendall2018multi}, the authors propose an automatic way to compute each loss weight $w_i$. For the MLM objective, the predicted token distribution is usually given as a Softmax:

\begin{equation}
   p(z|f_1(x, \Theta), \sigma_2) = \textrm{Softmax}\left(\frac{1}{\sigma_1^2}f_1(x,\Theta)\right)_z
\end{equation}
where $x$ is the input sequence, $f_1(x,\Theta)$ is the vector of logits outputted by the Language Head, $\Theta$ is the model parameters, and $\sigma_1$ is artificially added as a \textit{temperature} parameter. Then, after a few approximations, they get to write the following log-likelihood expression

\begin{equation}
    \log p(z|f_1(x,\Theta)) \propto \frac{1}{\sigma_1^2}\log\textrm{Softmax}\left(f_1(x,\Theta)\right)_z  - \log\sigma_1
\end{equation}
Here, one may recognize the cross-entropy loss with its weight $\frac{1}{\sigma_1^2}$ and a regularization term $\log\sigma_1$. Therefore, we take $L_1$ as the cross-entropy loss function, and we define $w_1 = \frac{1}{\sigma_1^2}$.

For the number prediction task, the authors predict the masked number using a Gaussian distribution:

\begin{equation}
    p\left(s|f_2(x, \Theta)\right) = \mathcal{N}\left(f_2(x, \Theta)|\sigma_2^2\right)
\end{equation}
where $f_2(x,\Theta)$ is Number Prediction Head's output, and $\mathcal{N}$ refers to the Gaussian probability distribution function and $\sigma_2$ its standard deviation. This leads to the log likelihood

\begin{equation}
    \log p(s|f_1(x,\Theta)) \propto -\frac{1}{2\sigma_2^2}\left(s-f_2(x,\Theta)\right)^2 - \log\sigma_2
\end{equation}
Then one can recognize the square error $\left(s-f_2(x,\Theta)\right)^2$ with its weight $\frac{1}{2\sigma_2^2}$ and a regularization term $\log\sigma_2$. Therefore, we take $L_2$ as the square error loss and we define $w_2 = \frac{1}{2\sigma_2^2}$.

Finally, when we sum up everything, we get the following loss function:

\begin{equation}
\label{eq:multi}
\begin{split}
    L(x, y_1, y_2) &= - \frac{1}{\sigma_1^2}\log\textrm{Softmax}\left(f_1(x,\Theta)\right)_{y_1} \\
    &\quad + \frac{1}{2\sigma_2^2}\left( y_2-f_2(x,\Theta)\right)^2 \\
    &\quad + \log\sigma_1 + \log\sigma_2\\
    &= \frac{1}{\sigma_1^2}L_1(x, y_1) + \frac{1}{2\sigma_2^2}L_2(x, y_2) \\
    &\quad + \log\sigma_1 + \log\sigma_2
\end{split}
\end{equation}

More precisely, the global objective function is
$$\mathcal{L}(\Theta) = \mathbb{E}_{\mathcal{D}}\left[L(x, y_1, y_2)\right]$$
with respect to $\Theta, \sigma_1, \sigma_2$ \cite{kendall2018multi}. The first-order conditions with respect to $\sigma_1$ and $\sigma_2$, give

\begin{equation}
\label{eq:1}
\sigma_1^2 =  2\mathbb{E}_{\mathcal{D}}\left[L_1(x, y_1)\right]    
\end{equation}

\begin{equation}
\label{eq:2}
\sigma_2^2 =  \mathbb{E}_{\mathcal{D}}\left[L_2(x, y_2)\right]    
\end{equation}

Since the exact numbers within a medical note are often not predictable from the context, and given the wide range of values present in the dataset (Figure ~\ref{fig:number distribution}), it is natural to assume that $\mathbb{E}_{\mathcal{D}}\left[L_2(x, y_2)\right]$ is significantly larger than $\mathbb{E}_{\mathcal{D}}\left[L_1(x, y_1)\right]$. Therefore by Equations ~\ref{eq:1} and ~\ref{eq:2}, $w_2$ is significantly smaller than $w_1$. Our experiments confirm this assumption, placing greater emphasis on contextual representation over magnitude-based representation. As a result, the magnitude objective is effectively suppressed during training, and the resulting classification performance degrades.

$\mathbb{E}_{\mathcal{D}}\left[L_2(x, y_2)\right]$ depends on the numbers distribution in $\mathcal{D}$ and more precisely on the range of the distribution. In order to lower the impact of this parameter, we substituted $L_2$ with its log-scaled version $\log (L_2)$. Additionally, we removed the regularization terms $\log\sigma_1$ and $\log\sigma_2$ and we adopted $w_1 = w_2 = \frac{1}{2}$

\begin{equation}
\label{eq:logloss}
\tilde{L}_2 (x, y_2) = \left( \log(y_2+1) - \log\left(f_2(x,\Theta)+1\right)\right)^2
\end{equation}

We therefore get the new multi-objective loss
\begin{equation}
    \tilde{L}(x, y_1, y_2) = \frac{1}{2}L_1(x, y_1) + \frac{1}{2}\tilde{L}_2(x, y_2)
\end{equation}

instead of ~\ref{eq:multi}.}

This substitution is motivated by the following considerations. Let $\theta \in \Theta$ represent a parameter that requires updating. For simplicity, we'll denote the partial derivative with respect to $\theta$ as $\partial_\theta$. We derive the two following gradients of $L_2$ and $\tilde{L}_2$ with respect to $\theta$:

\begin{equation}
    \begin{split}
    \partial_\theta L_2(x, y_2) &=\partial_\theta\left\{\left( y_2-f_2(x,\theta)\right)^2\right\}\\
    &= 2\partial_\theta f_2(x, \theta)\left(f_2(x, \theta) - y_2\right)
    \end{split}
\end{equation}

\begin{equation}
    \begin{split}
    \partial_\theta \tilde{L}_2(x, y_2) 
    &= \partial_\theta\left\{\left( \log(y_2+1) - \log\left(f_2(x,\theta)+1\right)\right)^2\right\}\\
    &= 2\frac{\partial_\theta f_2(x, \theta)}{f_2(x, \theta) + 1} \log\left(\frac{f_2(x, \theta)+1}{y_2+1}\right)
    \end{split}
\end{equation}

Now, the two gradients can be compared by examining their ratio.

\begin{equation}
\label{eq:ratio}
\begin{split}
    \left\lvert \frac{\partial_\theta \tilde{L}_2}{\partial_\theta L_2}\right\rvert
    = \Bigg\lvert &\frac{1}{\left(f_2(x, \theta) + 1\right)\left(f_2(x, \theta) - y_2\right)} \\
    &\times \log\left(\frac{f_2(x, \theta)+1}{y_2+1}\right)\Bigg\rvert
\end{split}
\end{equation}

Here, the term $\frac{1}{f_2(x, \theta) - y_2} \log\left(\frac{f_2(x, \theta)+1}{y_2+1}\right)$ is contingent on the relative comparison between the prediction $f_2(x, \theta)$ and $y_2$. However, the term $\frac{1}{f_2(x, \theta) + 1}$ exhibits a dependency on the magnitude of the prediction. In simpler terms, when high numbers (outliers) are predicted, $L_2$ will update the weights more strongly than $\tilde{L}_2$, potentially undermining the model's performance on the real data distribution. This characteristic underscores why $\tilde{L}_2$ is the more effective objective whenever the target values span a wide range of magnitudes.

\vspace{-5mm}

\rev{\subsection{Methodological Contributions}}\label{sec:novelty}

\rev{Although the proposed methodology builds on established components, its contribution lies in their formulation and integration for learning contextual and magnitude-aware representations of clinical numbers. Specifically, three methodological contributions distinguish our approach from the prior work summarised in Table \ref{tab:priorwork}.}

\rev{\textit{A failure mode of automatic loss weighting.} The scheme of \cite{kendall2018multi} is presented as general-purpose, but Equations \eqref{eq:1}--\eqref{eq:2} show its stationary points set $w_2 = 1/(2\sigma_2^2)$ to decay as $\mathbb{E}_{\mathcal{D}}[L_2]$ grows. In a corpus spanning $10^{-4}$ to $10^{5}$, the number head's squared error dwarfs the language head's cross-entropy, so $w_2 \to 0$ and the multi-objective model silently degenerates into the single-objective one, a degeneracy not previously reported since the method was validated on vision targets of comparable scale.}

\rev{\textit{A loss whose gradient is bounded by the predicted magnitude.} Replacing $L_2$ with $\tilde{L}_2$ (Equation \eqref{eq:logloss}) is not cosmetic: Equation \eqref{eq:ratio}'s gradient ratio carries the factor $1/(f_2(x,\theta)+1)$, so a single reading of $35{,}000$ can no longer dominate an update where most targets lie below $100$, letting the magnitude objective keep a fixed weight instead of being annealed away.}

\rev{\textit{A single encoder carrying both label semantics and magnitude.} LESA conditions attention on class-keyword embeddings while Xval injects magnitude into token embeddings before the attention layers; the two act on the same matrix at the same depth and had not previously been combined. Section \ref{sec:results} shows this is not redundant: the joint model departs significantly from a LESA-only encoder precisely on the \textit{Gradient} class, where context is ambiguous and only magnitude separates the categories.}

\rev{Algorithm \ref{alg:model2} states the resulting procedure. Stage 1 performs the multi-objective pre-finetuning on the unannotated corpus, and Stage 2 fine-tunes the encoder for token classification on the annotated subset.}

\begin{algorithm}[!t]
\ifhighlight\color{blue}\fi
\caption{Training of the proposed approach (CamemBERT-bio + LESA + Xval).}
\label{alg:model2}
\footnotesize
\begin{algorithmic}[1]
\Require Unannotated corpus $\mathcal{U}$, annotated corpus $\mathcal{A}$, class keyword sets $K_1,\dots,K_n$, pre-trained parameters $\Theta_0$
\Ensure Token classifier over the $n$ physiological classes
\Statex \textit{Stage 1: multi-objective pre-finetuning on $\mathcal{U}$}
\State $\boldsymbol{X}^{l}_j \gets \mathrm{mean}\left(\mathrm{Embed}(K_j)\right)$ for $j = 1,\dots,n$
\State $\Theta \gets \Theta_0$
\Repeat
  \For{each minibatch $B \subseteq \mathcal{U}$}
    \For{each note $x \in B$}
      \State $t_{masked} \gets$ replace every number in $x$ by \textit{NUM}
      \State mask one token; $y_1 \gets$ its vocabulary index
      \State $y_2 \gets$ its value if numeric, else $y_2 \gets 1$
      \State $h_{masked} \gets \mathrm{Embed}(t_{masked})$
      \State $h \gets$ scale \textit{NUM} rows of $h_{masked}$ by their values
      \State $\boldsymbol{A}^{l} \gets \mathrm{CrossAttention}(\boldsymbol{X}^{l}, h)$
      \State $C \gets \mathrm{norm}(\boldsymbol{A}^{l})^{T}\mathrm{norm}(\boldsymbol{A}^{l})$
      \State $S \gets \mathrm{SelfAttention}(h) + C$
      \State $f_1, f_2 \gets$ language head, number head applied to $S$
      \State $L_1 \gets \mathrm{CrossEntropy}(f_1, y_1)$
      \State $\tilde{L}_2 \gets \left(\log(y_2+1)-\log(f_2+1)\right)^{2}$
      \State $\tilde{L} \gets \tfrac{1}{2}L_1 + \tfrac{1}{2}\tilde{L}_2$
    \EndFor
    \State $\Theta \gets \mathrm{AdamW}\left(\Theta, \nabla_{\Theta}\,\mathrm{mean}_{B}\,\tilde{L}\right)$
  \EndFor
\Until{early stopping on validation loss}
\Statex \textit{Stage 2: supervised fine-tuning on $\mathcal{A}$}
\State replace the two pre-training heads by a token classification head
\Repeat
  \For{each minibatch $B \subseteq \mathcal{A}$}
    \State $\Theta \gets \mathrm{AdamW}\left(\Theta, \nabla_{\Theta}\,\mathrm{CrossEntropy}\right)$
  \EndFor
\Until{early stopping with patience $p$}
\State \Return fine-tuned token classifier
\end{algorithmic}
\end{algorithm}

\section{Experiments}\label{sec:experiments}
This section introduces the baseline models, outlines our experimental configuration, and presents the results we achieved across all classes.

\subsection{Our models}\label{our models}
We present two models that incorporate the diverse solutions proposed in this research.

\noindent \rev{\textbf{CamemBERT-bio + LESA (MLM-adapted):} This model uses the architecture presented in \cite{lompo2025mediumsizedtransformersmodelsrelevant}, but includes an additional adaptation stage on our corpus. Its training comprises the following two stages:}

\begin{enumerate}
    \item \rev{CamemBERT-bio augmented with Label Embedding for Self-Attention (LESA) is first adapted to the unannotated medical notes using the Masked Language Modeling (MLM) objective.}
    \item \rev{The resulting model is then fine-tuned for token classification using the annotated notes.} 
\end{enumerate}
\rev{Thus, the architecture is unchanged from \cite{lompo2025mediumsizedtransformersmodelsrelevant}; the distinction from the CamemBERT-bio + LESA baseline is the additional corpus-specific MLM adaptation stage. We use the compact label LESA--MLM in space-constrained tables.}

\noindent \rev{\textbf{CamemBERT-bio + LESA + Xval (proposed approach):} We extend CamemBERT-bio + LESA (MLM-adapted) with Xval to learn both contextual and magnitude-based representations of numbers. We use the compact label LESA--Xval in space-constrained tables.}
\vspace{-3mm}
\subsection{Baselines}

\rev{The baselines were selected so that each family in Table \ref{tab:priorwork} is represented by at least one system, isolating the contribution of every component rather than just the aggregate effect: ELMO for recurrent contextual encoders, DistilCamemBERT for knowledge distillation under small data, CamemBERT-bio as the unmodified domain-specific encoder, NumBERT for alternative numerical tokenization, CamemBERT-bio + Xval for magnitude without label semantics, CamemBERT-bio + LESA for label semantics without magnitude, and GPT-4 for large-scale instruction-tuned models. The two ablations, Xval-only and LESA-only, matter most, since comparing the proposed approach against each in isolation separates the effect of the multi-objective representation from either component alone.}

\begin{itemize}
  \item GPT-4: We selected 200 entries from our dataset and manually provided them to ChatGPT via few-shot prompting; the model's outputs were mapped to categories and aggregated to compute the evaluation metrics below.
  \item DistilCamemBERT \cite{delestre:hal-03674695}: reduces CamemBERT's parameters via Knowledge Distillation \cite{Bucila2006ModelC,hinton2015distilling}, a teacher-student training scheme; we used the pre-trained HuggingFace version \cite{wolf2019huggingface}.
  \item CamemBERT-bio: a French version of BioBERT \cite{Lee_2019}, introduced by \cite{touchent2023camembert} and trained on a large medical corpus.
  \item \rev{CamemBERT-bio + LESA \cite{lompo2025mediumsizedtransformersmodelsrelevant}: we retain the better-performing placeholder variant from that study; unlike CamemBERT-bio + LESA (MLM-adapted), it skips the corpus-specific MLM adaptation stage.}
  \item CamemBERT-bio + Xval \cite{golkar2023xval}: incorporates magnitude information into number embeddings; architecture imported from \href{https://github.com/PolymathicAI/xVal/blob/main/xval/numformer.py}{their repository} and trained under the protocol of Section \ref{the setup}.
  \item NumBERT \cite{zhang2020language}: substitutes each numerical instance with its scientific-notation (exponent/mantissa) representation; retrained on our corpus using \href{https://github.com/google-research/google-research/tree/master/numbert}{the reference implementation} under the settings of Table \ref{tab:hyperparams}.
  \item ELMO \cite{sarzynska2021detecting}: a bidirectional LSTM trained with a coupled language-model objective, effective in medical classification tasks; we used the \href{https://github.com/HIT-SCIR/ELMoForManyLangs?tab=readme-ov-file}{authors' implementation} directly, without further retraining.
\end{itemize}
\vspace{-3mm}

\subsection{Setup and Evaluation}\label{the setup}
\rev{The evaluation metric is the class-wise $F_1$ score, suitable for imbalanced datasets. Table \ref{tab:hyperparams} collects the full training configuration (optimizer, schedule, batch size, stopping rule, and so on) so the experiments can be reproduced without recovering individual values from the text; the same configuration was applied to every model in the comparison, so the differences reported in Section \ref{sec:results} are attributable to the representation and not to the training budget. For the downstream classification task we performed 10 runs with different random seeds and report the mean $F_1$ with standard deviation in Table \ref{results}. All methods used class-weighted cross-entropy to compensate for distribution imbalance, except the proposed approach, which performed better unweighted.}
\vspace{-3mm}

\begin{table}[!t]
\caption{\rev{Training configuration. Identical settings were applied to all models unless stated otherwise.}}
\label{tab:hyperparams}
\centering
\rev{%
\footnotesize
\begin{tabular}{@{}ll@{}}
\toprule
\textbf{Setting} & \textbf{Value} \\
\midrule
Base encoder & CamemBERT-bio, 110M parameters \\
Optimiser & AdamW \cite{loshchilov2017decoupled} \\
Learning rate & $3\times10^{-5}$ \\
Learning-rate schedule & Cosine, after warm-up \\
Warm-up & 100 epochs \\
Batch size & 32 \\
Maximum sequence length & 512 \\
MLM masking probability & 0.15 \\
Weight decay & 0.01 \\
Early stopping & Patience of 4 epochs \\
Independent runs & 10 random seeds \\
Train / validation / test & 70\% / 15\% / 15\% \\
Loss, baselines & Class-weighted cross-entropy \\
Loss, proposed approach & $\tfrac{1}{2}L_1 + \tfrac{1}{2}\tilde{L}_2$, unweighted \\
Hardware & Tesla V100, 32 GB \\
Training time, our models & $\approx 20$ hours \\
Training time, baselines & $\approx 10$ hours \\
\bottomrule
\end{tabular}}
\end{table} 
 
\section{Results and Discussions}\label{sec:results}
\vspace{-10mm}

\rev{\subsection{Overall performance}}

\rev{Table \ref{results} shows all models attain near-perfect $F_1$ on the Out-of-class category, expected given its extensive representation. We compute overall performance as the unweighted average $F_1$ across all classes, treating the remaining classes as equally important regardless of size, and report it per model in Table \ref{results global}.} \rev{This ranking shows that CamemBERT-bio + LESA (MLM-adapted) and CamemBERT-bio + LESA + Xval outperform the other baseline methods, except GPT-4.}

\begin{table}[!t]
\caption{Overall $F_1$ score per model, averaged over the eight classes.}
\label{results global}
\centering
\begin{tabular}{@{}lc@{}}
\toprule
\textbf{Model} & \textbf{Overall} $\mathbf{F_1}$ \textbf{score} \\
\midrule
GPT-4 & 0.95 \\
\addlinespace
ELMO & 0.72 \\
DistilCamemBERT & 0.74 \\
CamemBERT-bio & 0.69 \\
CamemBERT-bio + LESA & 0.83 \\
CamemBERT-bio + Xval & 0.80 \\
NumBERT & 0.77 \\
\addlinespace
\rev{Proposed LESA--MLM} & 0.83 \\
\rev{Proposed LESA--Xval} & \textbf{0.90} \\
\bottomrule
\end{tabular}
\end{table}

\subsection{Impact of pre-finetuning via MLM}
\rev{Table \ref{comparison} compares CamemBERT-bio and CamemBERT-bio + LESA before and after prefinetuning. Prefinetuning has no significant effect on CamemBERT-bio alone, expected since it was already trained on similar medical corpora and most of that knowledge is already encoded in its parameters. CamemBERT-bio + LESA, however, gains significantly, indicating that incorporating Label Embeddings requires fine-tuning to adjust the embedding layers so that the resulting latent space better discriminates the eight categories.}

\begin{table}[!t]
\caption{Impact of prefinetuning on the overall $F_1$ score.}
\label{comparison}
\centering
\begin{tabular}{@{}lcc@{}}
\toprule
\textbf{Model} & \textbf{Before} & \textbf{After} \\
\midrule
CamemBERT-bio & 0.69 & 0.70 \\
CamemBERT-bio + LESA & 0.79 & 0.83 \\
\bottomrule
\end{tabular}
\vspace{-3mm}
\end{table}

\subsection{Results of the multiple objective loss}
\begin{figure}
\centering
\includegraphics[scale=0.40]{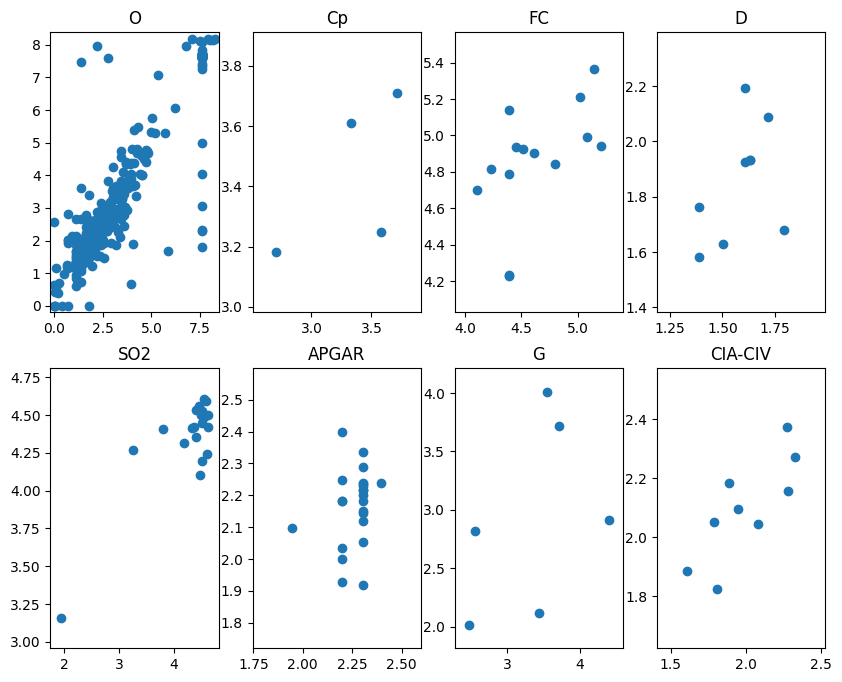}
  \caption{The numbers predicted by the proposed approach. The $x$ axis represents the ground truth values, and the $y$ axis represents the predicted values. Both axes are displayed on a logarithmic scale.\looseness=-1}
  \label{fig:predictions}
  \vspace{-6mm}
\end{figure}

\rev{The number prediction loss introduces intrinsic magnitude information into the model's embeddings, an effect significantly enhanced by Xval. We expect predictions close to, but not identical to, the ground truth, since exact reconstruction would indicate memorization rather than generalization; Figure \ref{fig:predictions} shows the proposed approach's predictions mostly aligning with ground truth, so we omit mean squared error as it is not a key metric here. Table \ref{results} breaks $F_1$ down by category (best score excluding GPT-4 in bold): the proposed approach outperforms every model but GPT-4 in nearly every class. All models, including GPT-4, struggle on \textit{ventricular Gradient}, which is confusable with aortic, aortic-ring, pulmonary, and inter-earcup gradients, a limitation of context-based representation since the surrounding context often provides insufficient cues and the gradient type is distinguished by magnitude alone (e.g., inter-earcup gradients are notably lower than ventricular ones). This magnitude-based distinction is exactly what gives the proposed approach its advantage in this category.}

\begin{table*}[!t]
\caption{$F_1$ score per class, reported as mean $\pm$ standard deviation over ten random seeds. For each class, the best score excluding GPT-4 is shown in bold. GPT-4 was evaluated once on a 200-entry subset and therefore carries no standard deviation.}
\label{results}
\centering
\footnotesize
\setlength{\tabcolsep}{4pt}
\begin{tabular}{@{}lcccccccc@{}}
\toprule
\textbf{Model} & \textbf{O} & \textbf{Cp} & \textbf{FC} & \textbf{D} & \textbf{SO2} & \textbf{APGAR} & \textbf{G} & \textbf{CIA/CIV} \\
\midrule
GPT-4 & 0.99 & 0.93 & 0.99 & 0.97 & 0.99 & 0.99 & 0.78 & 0.99 \\
\addlinespace
ELMO & 0.99 $\pm$ 0.00 & 0.76 $\pm$ 0.25 & 0.62 $\pm$ 0.18 & 0.68 $\pm$ 0.13 & 0.83 $\pm$ 0.15 & 0.87 $\pm$ 0.12 & 0.24 $\pm$ 0.14 & 0.81 $\pm$ 0.16 \\
DistilCamemBERT & 0.99 $\pm$ 0.00 & 0.80 $\pm$ 0.16 & 0.62 $\pm$ 0.09 & 0.69 $\pm$ 0.14 & 0.84 $\pm$ 0.05 & 0.95 $\pm$ 0.02 & 0.23 $\pm$ 0.02 & 0.81 $\pm$ 0.15 \\
CamemBERT-bio & 0.99 $\pm$ 0.00 & 0.78 $\pm$ 0.27 & 0.68 $\pm$ 0.07 & 0.52 $\pm$ 0.11 & 0.85 $\pm$ 0.05 & 0.93 $\pm$ 0.02 & 0.22 $\pm$ 0.02 & 0.80 $\pm$ 0.08 \\
CamemBERT-bio + LESA & 0.99 $\pm$ 0.00 & 0.85 $\pm$ 0.07 & 0.87 $\pm$ 0.04 & 0.84 $\pm$ 0.13 & 0.91 $\pm$ 0.02 & \textbf{0.99 $\pm$ 0.01} & 0.27 $\pm$ 0.01 & 0.91 $\pm$ 0.05 \\
CamemBERT-bio + Xval & 0.99 $\pm$ 0.00 & 0.85 $\pm$ 0.14 & 0.64 $\pm$ 0.07 & \textbf{0.92 $\pm$ 0.06} & 0.90 $\pm$ 0.03 & 0.95 $\pm$ 0.01 & 0.23 $\pm$ 0.15 & 0.91 $\pm$ 0.08 \\
NumBERT & 0.99 $\pm$ 0.00 & 0.78 $\pm$ 0.10 & 0.64 $\pm$ 0.04 & 0.67 $\pm$ 0.15 & 0.83 $\pm$ 0.07 & 0.90 $\pm$ 0.13 & 0.24 $\pm$ 0.10 & 0.82 $\pm$ 0.11 \\
\addlinespace
\rev{Proposed LESA--MLM} & 0.99 $\pm$ 0.00 & 0.85 $\pm$ 0.07 & 0.87 $\pm$ 0.04 & 0.84 $\pm$ 0.13 & 0.91 $\pm$ 0.02 & \textbf{0.99 $\pm$ 0.01} & 0.27 $\pm$ 0.01 & 0.91 $\pm$ 0.05 \\
\rev{Proposed LESA--Xval} & 0.99 $\pm$ 0.00 & \textbf{0.86 $\pm$ 0.07} & \textbf{0.89 $\pm$ 0.11} & 0.90 $\pm$ 0.08 & \textbf{0.93 $\pm$ 0.02} & 0.98 $\pm$ 0.03 & \textbf{0.61 $\pm$ 0.12} & \textbf{0.94 $\pm$ 0.06} \\
\bottomrule
\end{tabular}
\vspace{-3mm}
\end{table*}

\vspace{-5mm}

\rev{\subsection{Statistical significance of the comparisons}}

\rev{Mean scores alone do not establish that the differences in Table \ref{results} exceed run-to-run variability from random initialization. We therefore compare the proposed approach, CamemBERT-bio + LESA + Xval (labeled \textit{Proposed LESA--Xval} in Table \ref{results}), against each baseline with a two-sided Welch $t$-test, including against \textit{LESA--MLM}, the ablation of the proposed approach without the number-prediction objective, applied independently per class, which does not assume equal variances and so remains valid under the unequal per-class standard deviations observed in Table \ref{results} (e.g., $0.01$ to $0.27$, and a variance ratio of $144$ on the \textit{Gradient} class against LESA--MLM). With $\bar{x}_1,s_1,n_1$ and $\bar{x}_2,s_2,n_2$ the mean, sample standard deviation and run count of the proposed approach and a baseline on a given class, and $u_k = s_k^2/n_k$,
\begin{equation}
\label{eq:welch-t}
t = \frac{\bar{x}_1 - \bar{x}_2}{\sqrt{u_1 + u_2}}, \qquad
\nu = \frac{\left(u_1 + u_2\right)^{2}}
           {\dfrac{u_1^{2}}{n_1 - 1} + \dfrac{u_2^{2}}{n_2 - 1}}
\end{equation}
where $\nu$ is the Welch--Satterthwaite degrees of freedom, and the two-sided $p$-value is $p = 2[1-F_\nu(|t|)]$ with $F_\nu$ the Student CDF. With $n_1=n_2=10$, this statistic is numerically identical to the pooled Student test but attributes fewer degrees of freedom whenever the two variances differ ($\nu \le 18$), which is the conservative choice under the dispersion observed here \cite{dewinter2013student, delacre2017welch}; every $p$-value we report is therefore an upper bound on the one a pooled test would give. Ten runs per model is enough for the $t$-test to be admissible \cite{dewinter2013student}. The \textit{Out-of-class} category is excluded because every model attains $0.99\pm0.00$, leaving the statistic undefined. Results are in Table \ref{tab:ttest}.}

\begin{table*}[!t]
\caption{\rev{Two-sided Welch $t$-test comparing the proposed approach (CamemBERT-bio + LESA + Xval) against each baseline, per class, over ten seeds ($p$-values). The \textit{LESA--MLM} row is the ablation of the proposed approach without the number-prediction objective. A positive $\Delta$ favors the proposed approach. Significance is coded as *** $p<0.001$, ** $p<0.01$, * $p<0.05$, ns $p\geq0.05$. The \textit{Out-of-class} category is omitted; all models score $0.99\pm0.00$.}}
\label{tab:ttest}
\centering
\rev{%
\footnotesize
\begin{tabular}{@{}lccccccc@{}}
\toprule
\multicolumn{8}{c}{\textbf{vs. Proposed LESA--Xval (CamemBERT-bio + LESA + Xval)}} \\
\midrule
\textbf{Baseline} & \textbf{Cp} & \textbf{FC} & \textbf{D} & \textbf{SO2} & \textbf{APGAR} & \textbf{G} & \textbf{CIA/CIV} \\
\midrule
ELMO               & 0.250\,ns & 0.001\,**  & $<$0.001\,*** & 0.065\,ns & 0.018\,*  & $<$0.001\,*** & 0.034\,*  \\
DistilCamemBERT    & 0.298\,ns & $<$0.001\,*** & $<$0.001\,*** & $<$0.001\,*** & 0.018\,*  & $<$0.001\,*** & 0.026\,*  \\
CamemBERT-bio      & 0.385\,ns & $<$0.001\,*** & $<$0.001\,*** & $<$0.001\,*** & $<$0.001\,*** & $<$0.001\,*** & $<$0.001\,*** \\
CamemBERT-bio + LESA & 0.753\,ns & 0.599\,ns & 0.233\,ns & 0.038\,*  & 0.339\,ns & $<$0.001\,*** & 0.241\,ns \\
CamemBERT-bio + Xval & 0.843\,ns & $<$0.001\,*** & 0.536\,ns & 0.018\,*  & 0.012\,*  & $<$0.001\,*** & 0.356\,ns \\
NumBERT            & 0.055\,ns & $<$0.001\,*** & $<$0.001\,*** & 0.001\,**  & 0.087\,ns & $<$0.001\,*** & 0.009\,**  \\
\addlinespace
\rev{LESA--MLM (ablation)} & 0.753\,ns & 0.599\,ns & 0.233\,ns & 0.038\,*  & 0.339\,ns & $<$0.001\,*** & 0.241\,ns \\
\bottomrule
\end{tabular}}
\vspace{-3mm}
\end{table*}

\rev{The test sharpens the interpretation of the results in three ways. Against the four representations that carry neither label semantics nor magnitude, namely ELMO, DistilCamemBERT, CamemBERT-bio and NumBERT, the proposed approach is significantly better in five to six of the seven classes, so the aggregate gap in Table \ref{results global} is not an artifact of seed variance. Against CamemBERT-bio + Xval, which supplies magnitude but no label semantics, the advantage narrows to four classes and disappears on \textit{D} and \textit{CIA/CIV}, the two categories whose values are already well separated in magnitude.}

\rev{The third comparison is the informative one. Against CamemBERT-bio + LESA, the proposed approach is significantly better in only two of seven classes, \textit{SO2} at $p=0.038$ and \textit{Gradient} at $p<0.001$. On \textit{Cp}, \textit{FC}, \textit{D}, \textit{APGAR} and \textit{CIA/CIV} the two models are statistically indistinguishable. We read this as a targeted result, not a weak one: adding a magnitude objective to an encoder that already carries label semantics does not improve classes that context alone resolves, and it should not be expected to. The improvement is concentrated where the premise of the method applies, on the class whose members are separable by magnitude but not by their surroundings. For \textit{Gradient} the difference is $+0.34$ $F_1$ with $t=8.93$, which is the largest effect in the table by a wide margin and is what carries the $0.07$ difference in the overall score reported in Table \ref{results global}.}

\rev{Two caveats apply. Equation \eqref{eq:welch-t} treats the runs of the two models as independent, whereas each is evaluated on the same fixed test split; a paired test over matched seed initializations would be more powerful, but we report the unpaired form as it needs no shared seeds across models. Also, no correction for multiple comparisons was applied, so isolated results with $p$ between $0.01$ and $0.05$ should be read as indicative; the conclusions above rest on the \textit{Gradient} column, where the effect survives any reasonable correction.}

\begin{figure}
\centering
\includegraphics[width=0.9\linewidth]{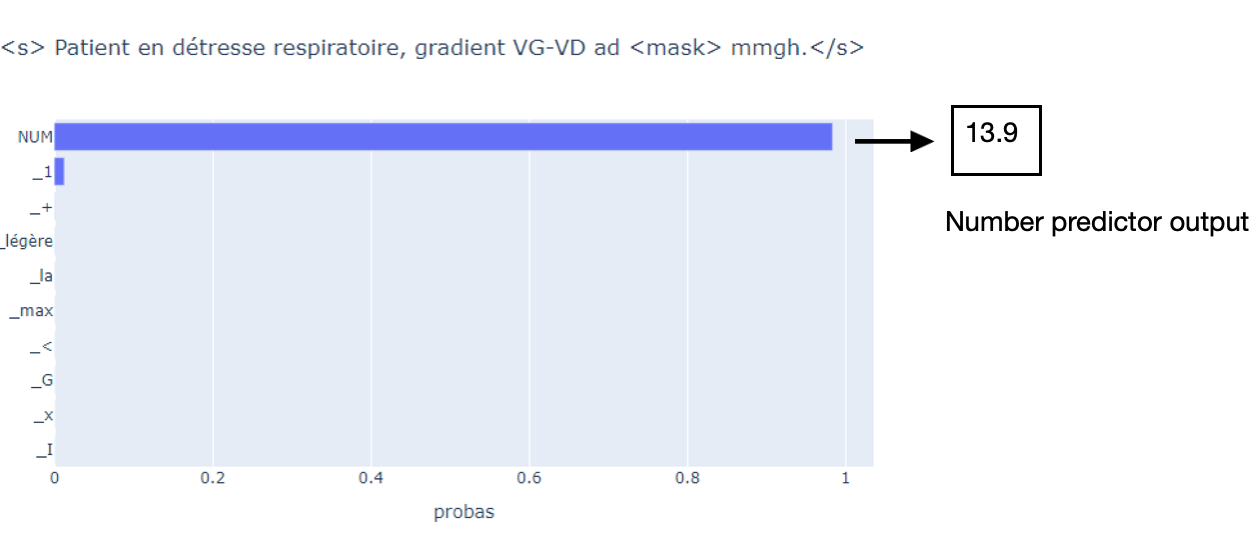}
  \caption{Output of a completion task by the proposed approach. The input sentence is \textit{``Patient en détresse respiratoire, gradient VG-VD ad $\langle$ mask $\rangle$ mmgh."}. The figure contains the possible words to fill the mask with their corresponding probabilities. Here, the model successfully predicts a number, which is given by the number prediction head.}
  \label{completion with Model 2}
  \vspace{-3mm}
\end{figure}

\begin{figure}
\centering
\includegraphics[width=0.9\linewidth]{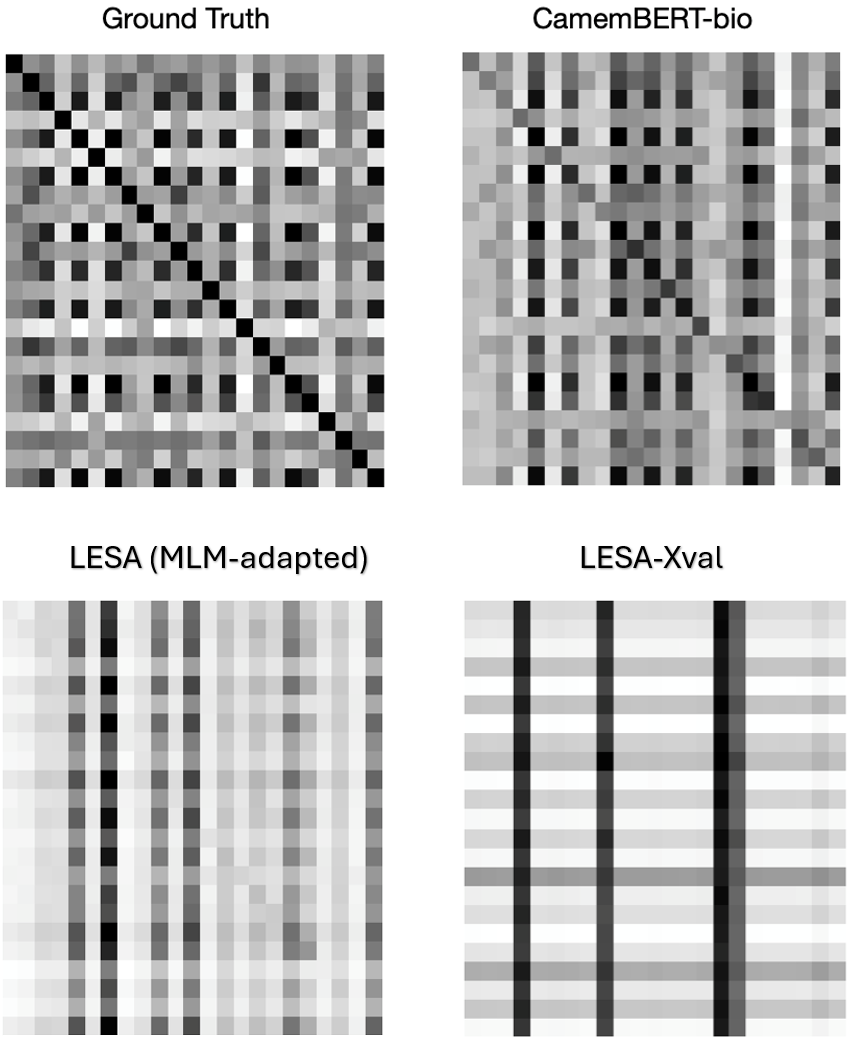}
  \caption{\rev{Effect of multi-objective representation on word embeddings. From left to right, the cosine-similarity matrices compare CamemBERT-bio embeddings with those from CamemBERT-bio itself, fine-tuned CamemBERT-bio, CamemBERT-bio + LESA (MLM-adapted), and CamemBERT-bio + LESA + Xval. Darker cells indicate a higher pairwise similarity score between terms.}}
  \label{comparison with Model 2}
  \vspace{-5mm}
\end{figure}

\rev{Subjecting the proposed approach to the same sentence completion test as Section \ref{tokenization problem} gives Figure \ref{completion with Model 2}: unlike CamemBERT-bio, it clearly recognizes that a numerical value is expected and its Number Head predicts $13.9$, within the correct range for this gradient type. We attribute this to placeholder replacement clarifying the unstructured text, and to LESA and Xval improving the predicted value range. To further illustrate this, we measure the cosine similarity between the embeddings of twenty frequently occurring medical terms produced by CamemBERT-bio (our reference) and by three derived models, CamemBERT-bio adapted via MLM, CamemBERT-bio + LESA (MLM-adapted), and the proposed approach, following \cite{steck2024cosine}, which shows cosine similarity is meaningful for models trained with respect to that metric, a condition our layer-normalized models satisfy.}

Figure \ref{comparison with Model 2} shows, for each model, the grayscale cosine-similarity matrix $M$ between the reference and that model's embeddings of the same terms; darker cells indicate higher similarity. \rev{From left to right, the matrices progressively lighten, indicating growing divergence from the reference embeddings. The second matrix shows CamemBERT-bio embeddings largely unchanged after fine-tuning alone, corroborating Table \ref{comparison}'s finding that fine-tuning adds no new knowledge. The third shows a notable shift once LESA and MLM adaptation are added, consistent with the corpus-specific adaptation modifying the embedding space after label information is introduced. The last matrix shows the largest deviation, supporting the hypothesis that the multi-objective loss transforms the embedding space more substantially than LESA alone, without a magnitude signal.}

\subsection{GPT-4 versus the proposed approach}\label{sec:gpt vs model 2}

\rev{GPT-4's overall $F_1$ exceeds the proposed approach by 0.05, a modest margin given that the proposed approach carries 110 million parameters against GPT-4's multi-trillion-parameter scale. A failure-case analysis by \cite{lompo2025mediumsizedtransformersmodelsrelevant} found GPT-4's recurring misclassifications concentrated in the Contractility (Cp) and Gradient (G) categories, from ambiguous abbreviations and confusion between gradient types, respectively; Table \ref{results} shows the proposed approach struggles with the same two categories, but narrows the CamemBERT-bio-to-GPT-4 gap in the G category from $0.52$ to $0.17$ $F_1$.}

\rev{Practically, the proposed approach is also cheaper to run and easier to integrate: unlike GPT-4 or LLaMA, it can be deployed locally at competitive performance without the associated compute cost, removing the dependence on external cloud services and keeping patient data on hospital infrastructure.}

\section{Limitations and Application Perspectives}\label{sec:limits}
\rev{The main limitation of this study is the choice of CamemBERT-bio over large-scale LLMs. This was a deliberate trade-off, not an oversight: with only 1,072 annotated samples and training GPUs shared concurrently with other hospital workloads, a multi-trillion-parameter model was not a realistic option, even though GPT-4 attains higher overall $F_1$. We also recognize that our experiments target a single task on a single-site dataset, though we expect the methodology to transfer to other settings sharing the same limited-data, shared-compute profile.}

\rev{The scope of the baseline comparison is a further limitation. Table \ref{tab:priorwork}'s families are covered, but instruction-tuned open-weight models (e.g., LLaMA) and newer French clinical encoders were not evaluated; an open-weight model in particular would be deployable under the same hospital constraints as ours and is the most immediate direction for further work. GPT-4 was the only large-scale system to which the annotated data could be exposed under this corpus's governance. Separately, the significance analysis of Table \ref{tab:ttest} rests on ten runs per model, which supports the large \textit{Gradient} effect but gives limited power for one- or two-point $F_1$ differences; several comparisons against CamemBERT-bio + LESA are therefore inconclusive rather than negative.}

\rev{Two model-side developments define the next iteration of this pipeline: the second release of CamemBERT-bio as the backbone \cite{touchent2023camembert}, to which LESA and the log-scaled magnitude loss transfer without modification, and inserting a causal language-modeling phase into Stage 1's continued pretraining \cite{touchent2026causal}, motivated by the uneven benefit MLM pre-finetuning currently gives CamemBERT-bio ($0.69\to0.70$) versus CamemBERT-bio + LESA ($0.79\to0.83$ in Table \ref{comparison}).}

The CDSS is still under development. Our next step is to integrate this NLP model with previously developed algorithms for heart failure detection \cite{le2022detecting, le2023adaptation, le2023improving}, hypoxemia detection \cite{sauthier2021estimated}, and chest X-ray analysis \cite{zaglam2014computer, yahyatabar2020dense}, toward deployment within the PICU for early ARDS diagnosis, followed by a prospective evaluation once integrated into the PICU's e-Medical infrastructure.

\section{Conclusion}\label{sec:conclusion}

\rev{This work addressed numerical understanding in French pediatric ICU notes under two binding constraints, a small annotated corpus and hospital compute shared with other clinical workloads, that rule out large-scale LLMs. We showed that CamemBERT-bio + LESA benefits from corpus-specific MLM prefinetuning, an improvement not seen with CamemBERT-bio alone, and combined it with Xval into a multi-objective representation carrying both contextual and magnitude information. The resulting model, at 110 million parameters, approaches GPT-4's performance and significantly outperforms every other baseline, offering a practical, locally deployable alternative to large-scale LLMs for medical note processing under limited data and shared compute.}

\section*{Acknowledgment}

\rev{This work was supported in part by the Natural Sciences and Engineering Research Council (NSERC), in part by the Institut de Valorisation des données de l’Université de Montréal (IVADO), in part by the Fonds de la recherche en sante du Quebec (FRQS). Data and reproducible codes are available upon request from Prof. Philippe Jouvet, M.D., PhD.}

\balance   
\bibliographystyle{IEEEtran}
\bibliography{Bibliography}

\begin{thebibliography}{10}
\providecommand{\url}[1]{#1}
\csname url@rmstyle\endcsname
\providecommand{\newblock}{\relax}
\providecommand{\bibinfo}[2]{#2}
\providecommand\BIBentrySTDinterwordspacing{\spaceskip=0pt\relax}
\providecommand\BIBentryALTinterwordstretchfactor{4}
\providecommand\BIBentryALTinterwordspacing{\spaceskip=\fontdimen2\font plus
\BIBentryALTinterwordstretchfactor\fontdimen3\font minus
  \fontdimen4\font\relax}
\providecommand\BIBforeignlanguage[2]{{%
\expandafter\ifx\csname l@#1\endcsname\relax
\typeout{** WARNING: IEEEtran.bst: No hyphenation pattern has been}%
\typeout{** loaded for the language `#1'. Using the pattern for}%
\typeout{** the default language instead.}%
\else
\language=\csname l@#1\endcsname
\fi
#2}}

\bibitem{touchent2023camembert}
R.~Touchent, L.~Romary, and E.~de~La~Clergerie, ``Camembert-bio: a tasty french
  language model better for your health,'' \emph{arXiv preprint
  arXiv:2306.15550}, 2023.

\bibitem{lompo2025mediumsizedtransformersmodelsrelevant}
\BIBentryALTinterwordspacing
B.~A. Lompo, T.-D. Le, P.~Jouvet, and R.~Noumeir, ``Are medium-sized
  transformers models still relevant for medical records processing?'' 2025.
  [Online]. Available: \url{https://arxiv.org/abs/2404.10171}
\BIBentrySTDinterwordspacing

\bibitem{golkar2023xval}
S.~Golkar, M.~Pettee, M.~Eickenberg, A.~Bietti, M.~Cranmer, G.~Krawezik,
  F.~Lanusse, M.~McCabe, R.~Ohana, L.~Parker, B.~R{\'e}galdo-Saint~Blancard,
  T.~Tesileanu, K.~Cho, and S.~Ho, ``xval: A continuous number encoding for
  large language models,'' \emph{arXiv preprint arXiv:2310.02989}, 2023.

\bibitem{kendall2018multi}
A.~Kendall, Y.~Gal, and R.~Cipolla, ``Multi-task learning using uncertainty to
  weigh losses for scene geometry and semantics,'' in \emph{Proceedings of the
  IEEE conference on computer vision and pattern recognition}, 2018, pp.
  7482--7491.

\bibitem{wallace2019nlp}
E.~Wallace, Y.~Wang, S.~Li, S.~Singh, and M.~Gardner, ``Do nlp models know
  numbers? probing numeracy in embeddings,'' \emph{arXiv preprint
  arXiv:1909.07940}, 2019.

\bibitem{cui2019regular}
M.~Cui, R.~Bai, Z.~Lu, X.~Li, U.~Aickelin, and P.~Ge, ``Regular expression
  based medical text classification using constructive heuristic approach,''
  \emph{IEEE Access}, vol.~7, pp. 147\,892--147\,904, 2019.

\bibitem{chen2023improving}
C.-C. Chen, H.~Takamura, I.~Kobayashi, and Y.~Miyao, ``Improving numeracy by
  input reframing and quantitative pre-finetuning task,'' in \emph{Findings of
  the Association for Computational Linguistics: EACL 2023}, 2023, pp. 69--77.

\bibitem{charton2021linear}
F.~Charton, ``Linear algebra with transformers,'' \emph{arXiv preprint
  arXiv:2112.01898}, 2021.

\bibitem{Loukas_2022}
\BIBentryALTinterwordspacing
L.~Loukas, M.~Fergadiotis, I.~Chalkidis, E.~Spyropoulou, P.~Malakasiotis,
  I.~Androutsopoulos, and G.~Paliouras, ``Finer: Financial numeric entity
  recognition for xbrl tagging,'' in \emph{Proceedings of the 60th Annual
  Meeting of the Association for Computational Linguistics (Volume 1: Long
  Papers)}.\hskip 1em plus 0.5em minus 0.4em\relax Association for
  Computational Linguistics, 2022. [Online]. Available:
  \url{http://dx.doi.org/10.18653/v1/2022.acl-long.303}
\BIBentrySTDinterwordspacing

\bibitem{Nguyen2018EmbNumSL}
\BIBentryALTinterwordspacing
P.~Nguyen, K.~Nguyen, R.~Ichise, and H.~Takeda, ``Embnum: Semantic labeling for
  numerical values with deep metric learning,'' \emph{ArXiv}, vol.
  abs/1807.01367, 2018. [Online]. Available:
  \url{https://api.semanticscholar.org/CorpusID:49567815}
\BIBentrySTDinterwordspacing

\bibitem{cheng2024potential}
N.~Cheng, Z.~Yan, Z.~Wang, Z.~Li, J.~Yu, Z.~Zheng, K.~Tu, J.~Xu, and W.~Han,
  ``Potential and limitations of llms in capturing structured semantics: A case
  study on srl,'' in \emph{International Conference on Intelligent
  Computing}.\hskip 1em plus 0.5em minus 0.4em\relax Springer, 2024, pp.
  50--61.

\bibitem{shah2024accuracy}
S.~V. Shah, ``Accuracy, consistency, and hallucination of large language models
  when analyzing unstructured clinical notes in electronic medical records,''
  \emph{JAMA Network Open}, vol.~7, no.~8, pp. e2\,425\,953--e2\,425\,953,
  2024.

\bibitem{ullah2024challenges}
E.~Ullah, A.~Parwani, M.~M. Baig, and R.~Singh, ``Challenges and barriers of
  using large language models (llm) such as chatgpt for diagnostic medicine
  with a focus on digital pathology--a recent scoping review,''
  \emph{Diagnostic pathology}, vol.~19, no.~1, p.~43, 2024.

\bibitem{Li2018}
\BIBentryALTinterwordspacing
Y.~Li, L.~Yao, C.~Mao, A.~Srivastava, X.~Jiang, and Y.~Luo, ``Early prediction
  of acute kidney injury in critical care setting using clinical notes,'' in
  \emph{2018 {IEEE} International Conference on Bioinformatics and Biomedicine
  ({BIBM})}.\hskip 1em plus 0.5em minus 0.4em\relax {IEEE}, Dec. 2018.
  [Online]. Available: \url{https://doi.org/10.1109/bibm.2018.8621574}
\BIBentrySTDinterwordspacing

\bibitem{ezen2020comparison}
A.~Ezen-Can, ``A comparison of lstm and bert for small corpus,'' \emph{arXiv
  preprint arXiv:2009.05451}, 2020.

\bibitem{mascio2020comparative}
A.~Mascio, Z.~Kraljevic, D.~Bean, R.~Dobson, R.~Stewart, R.~Bendayan, and
  A.~Roberts, ``Comparative analysis of text classification approaches in
  electronic health records,'' \emph{arXiv preprint arXiv:2005.06624}, 2020.

\bibitem{Bucila2006ModelC}
C.~Bucila, R.~Caruana, and A.~Niculescu-Mizil, ``Model compression,'' in
  \emph{Knowledge Discovery and Data Mining}, 2006.

\bibitem{Sanh2019}
\BIBentryALTinterwordspacing
V.~Sanh, L.~Debut, J.~Chaumond, and T.~Wolf, ``Distilbert, a distilled version
  of bert: smaller, faster, cheaper and lighter,'' 2019. [Online]. Available:
  \url{https://arxiv.org/abs/1910.01108}
\BIBentrySTDinterwordspacing

\bibitem{delestre2022distilcamembert}
C.~Delestre and A.~Amar, ``Distilcamembert: a distillation of the french model
  camembert,'' \emph{arXiv preprint arXiv:2205.11111}, 2022.

\bibitem{zhang2020language}
X.~Zhang, D.~Ramachandran, I.~Tenney, Y.~Elazar, and D.~Roth, ``Do language
  embeddings capture scales?'' \emph{arXiv preprint arXiv:2010.05345}, 2020.

\bibitem{hu2024improving}
Y.~Hu, Q.~Chen, J.~Du, X.~Peng, V.~K. Keloth, X.~Zuo, Y.~Zhou, Z.~Li, X.~Jiang,
  Z.~Lu, K.~Roberts, and H.~Xu, ``Improving large language models for clinical
  named entity recognition via prompt engineering,'' \emph{Journal of the
  American Medical Informatics Association}, vol.~31, no.~9, pp. 1812--1820,
  2024.

\bibitem{wu2016google}
Y.~Wu, M.~Schuster, Z.~Chen, Q.~V. Le, M.~Norouzi, W.~Macherey, M.~Krikun,
  Y.~Cao, Q.~Gao, K.~Macherey, J.~Klingner, A.~Shah, M.~Johnson, X.~Liu,
  {\L}.~Kaiser, S.~Gouws, Y.~Kato, T.~Kudo, H.~Kazawa, K.~Stevens, G.~Kurian,
  N.~Patil, W.~Wang, C.~Young, J.~Smith, J.~Riesa, A.~Rudnick, O.~Vinyals,
  G.~Corrado, M.~Hughes, and J.~Dean, ``Google's neural machine translation
  system: Bridging the gap between human and machine translation,'' \emph{arXiv
  preprint arXiv:1609.08144}, 2016.

\bibitem{kudo-richardson-2018-sentencepiece}
\BIBentryALTinterwordspacing
T.~Kudo and J.~Richardson, ``{S}entence{P}iece: A simple and language
  independent subword tokenizer and detokenizer for neural text processing,''
  in \emph{Proceedings of the 2018 Conference on Empirical Methods in Natural
  Language Processing: System Demonstrations}.\hskip 1em plus 0.5em minus
  0.4em\relax Brussels, Belgium: Association for Computational Linguistics,
  Nov. 2018, pp. 66--71. [Online]. Available:
  \url{https://aclanthology.org/D18-2012}
\BIBentrySTDinterwordspacing

\bibitem{clark2019does}
K.~Clark, U.~Khandelwal, O.~Levy, and C.~D. Manning, ``What does bert look at?
  an analysis of bert's attention,'' \emph{arXiv preprint arXiv:1906.04341},
  2019.

\bibitem{collobert2008unified}
R.~Collobert and J.~Weston, ``A unified architecture for natural language
  processing: Deep neural networks with multitask learning,'' in
  \emph{Proceedings of the 25th international conference on Machine learning},
  2008, pp. 160--167.

\bibitem{delestre:hal-03674695}
\BIBentryALTinterwordspacing
C.~Delestre and A.~Amar, ``{DistilCamemBERT : une distillation du mod{\`e}le
  fran{\c c}ais CamemBERT},'' in \emph{{CAp (Conf{\'e}rence sur l'Apprentissage
  automatique)}}, Vannes, France, July 2022. [Online]. Available:
  \url{https://hal.archives-ouvertes.fr/hal-03674695}
\BIBentrySTDinterwordspacing

\bibitem{hinton2015distilling}
G.~Hinton, O.~Vinyals, and J.~Dean, ``Distilling the knowledge in a neural
  network,'' \emph{arXiv preprint arXiv:1503.02531}, 2015.

\bibitem{wolf2019huggingface}
T.~Wolf, L.~Debut, V.~Sanh, J.~Chaumond, C.~Delangue, A.~Moi, P.~Cistac,
  T.~Rault, R.~Louf, M.~Funtowicz, J.~Davison, S.~Shleifer, P.~von Platen,
  C.~Ma, Y.~Jernite, J.~Plu, C.~Xu, T.~Le~Scao, S.~Gugger, M.~Drame, Q.~Lhoest,
  and A.~M. Rush, ``Huggingface's transformers: State-of-the-art natural
  language processing,'' \emph{arXiv preprint arXiv:1910.03771}, 2019.

\bibitem{Lee_2019}
\BIBentryALTinterwordspacing
J.~Lee, W.~Yoon, S.~Kim, D.~Kim, S.~Kim, C.~H. So, and J.~Kang, ``{BioBERT}: a
  pre-trained biomedical language representation model for biomedical text
  mining,'' \emph{Bioinformatics}, vol.~36, no.~4, pp. 1234--1240, sep 2019.
  [Online]. Available: \url{https://doi.org/10.1093%2Fbioinformatics%2Fbtz682}
\BIBentrySTDinterwordspacing

\bibitem{sarzynska2021detecting}
J.~Sarzynska-Wawer, A.~Wawer, A.~Pawlak, J.~Szymanowska, I.~Stefaniak,
  M.~Jarkiewicz, and L.~Okruszek, ``Detecting formal thought disorder by deep
  contextualized word representations,'' \emph{Psychiatry Research}, vol. 304,
  p. 114135, 2021.

\bibitem{loshchilov2017decoupled}
I.~Loshchilov and F.~Hutter, ``Decoupled weight decay regularization,''
  \emph{arXiv preprint arXiv:1711.05101}, 2017.

\bibitem{dewinter2013student}
J.~C.~F. de~Winter, ``Using the student's t-test with extremely small sample
  sizes,'' \emph{Practical Assessment, Research \& Evaluation}, vol.~18,
  no.~10, pp. 1--12, 2013.

\bibitem{delacre2017welch}
M.~Delacre, D.~Lakens, and C.~Leys, ``Why psychologists should by default use
  welch's t-test instead of student's t-test,'' \emph{International Review of
  Social Psychology}, vol.~30, no.~1, pp. 92--101, 2017.

\bibitem{steck2024cosine}
H.~Steck, C.~Ekanadham, and N.~Kallus, ``Is cosine-similarity of embeddings
  really about similarity?'' in \emph{Companion Proceedings of the ACM on Web
  Conference 2024}, 2024, pp. 887--890.

\bibitem{touchent2026causal}
R.~Touchent and E.~de~la Clergerie, ``A causal language modeling detour
  improves encoder continued pretraining,'' \emph{arXiv preprint
  arXiv:2605.12438}, 2026.

\bibitem{le2022detecting}
T.-D. Le, R.~Noumeir, J.~Rambaud, G.~Sans, and P.~Jouvet, ``Detecting of a
  patient's condition from clinical narratives using natural language
  representation,'' \emph{IEEE Open Journal of Engineering in Medicine and
  Biology}, vol.~3, pp. 142--149, 2022.

\bibitem{le2023adaptation}
------, ``Adaptation of autoencoder for sparsity reduction from clinical notes
  representation learning,'' \emph{IEEE Journal of Translational Engineering in
  Health and Medicine}, vol.~11, pp. 469--478, 2023.

\bibitem{le2023improving}
T.-D. Le, P.~Jouvet, and R.~Noumeir, ``Improving transformer performance for
  french clinical notes classification using mixture of experts on a limited
  dataset,'' \emph{IEEE Journal of Translational Engineering in Health and
  Medicine}, vol.~13, pp. 261--274, 2025.

\bibitem{sauthier2021estimated}
M.~Sauthier, G.~Tuli, P.~A. Jouvet, J.~S. Brownstein, and A.~G. Randolph,
  ``Estimated pao2: A continuous and noninvasive method to estimate pao2 and
  oxygenation index,'' \emph{Critical care explorations}, vol.~3, no.~10, p.
  e0546, 2021.

\bibitem{zaglam2014computer}
N.~Zaglam, P.~Jouvet, O.~Flechelles, G.~Emeriaud, and F.~Cheriet,
  ``Computer-aided diagnosis system for the acute respiratory distress syndrome
  from chest radiographs,'' \emph{Computers in biology and medicine}, vol.~52,
  pp. 41--48, 2014.

\bibitem{yahyatabar2020dense}
M.~Yahyatabar, P.~Jouvet, and F.~Cheriet, ``Dense-unet: a light model for lung
  fields segmentation in chest x-ray images,'' in \emph{2020 42nd Annual
  International Conference of the IEEE Engineering in Medicine \& Biology
  Society (EMBC)}.\hskip 1em plus 0.5em minus 0.4em\relax IEEE, 2020, pp.
  1242--1245.

\end{thebibliography}

\end{document}